\documentclass[preprintnumbers,superscriptaddress,amsmath,amssymb,pre,aps,10pt]{revtex4-1}

\usepackage{graphicx}
\usepackage{dcolumn}
\usepackage{bm}
\usepackage{color}
\usepackage{multirow}
\usepackage{listings}
\usepackage{CJK}

\begin{document}
\begin{CJK*}{GBK}{Song} 


\title{Effects of polynomial trends on detrending moving average analysis}

\author{Ying-Hui Shao}
 \affiliation{School of Business, East China University of Science and Technology, Shanghai 200237, China} %
 \affiliation{Research Center for Econophysics, East China University of Science and Technology, Shanghai 200237, China} %

\author{Gao-Feng Gu}
 \affiliation{School of Business, East China University of Science and Technology, Shanghai 200237, China} %
 \affiliation{Research Center for Econophysics, East China University of Science and Technology, Shanghai 200237, China} %

 \author{Zhi-Qiang Jiang}
 \affiliation{School of Business, East China University of Science and Technology, Shanghai 200237, China} %
 \affiliation{Research Center for Econophysics, East China University of Science and Technology, Shanghai 200237, China} %

 \author{Wei-Xing Zhou}
 \email{wxzhou@ecust.edu.cn}
 \affiliation{School of Business, East China University of Science and Technology, Shanghai 200237, China} %
 \affiliation{Research Center for Econophysics, East China University of Science and Technology, Shanghai 200237, China} %
 \affiliation{School of Science, East China University of Science and Technology, Shanghai 200237, China} %

\date{\today}

\begin{abstract}
  The detrending moving average (DMA) algorithm is one of the best performing methods to quantify the long-term correlations in nonstationary time series. Many long-term correlated time series in real systems contain various trends. We investigate the effects of polynomial trends on the scaling behaviors and the performances of three widely used DMA methods including backward algorithm (BDMA), centered algorithm (CDMA) and forward algorithm (FDMA). We derive a general framework for polynomial trends and obtain analytical results for constant shifts and linear trends. We find that the behavior of the CDMA method is not influenced by constant shifts. In contrast, linear trends cause a crossover in the CDMA fluctuation functions. We also find that constant shifts and linear trends cause crossovers in the fluctuation functions obtained from the BDMA and FDMA methods. When a crossover exists, the scaling behavior at small scales comes from the intrinsic time series while that at large scales is dominated by the constant shifts or linear trends. We also derive analytically the expressions of crossover scales and show that the crossover scale depends on the strength of the polynomial trend, the Hurst index, and in some cases (linear trends for BDMA and FDMA) the length of the time series. In all cases, the BDMA and the FDMA behave almost the same under the influence of constant shifts or linear trends. Extensive numerical experiments confirm excellently the analytical derivations. We conclude that the CDMA method outperforms the BDMA and FDMA methods in the presence of polynomial trends.

  {\textit{Keywords}}: Fractal Analysis; Detrending Moving Average (DMA); Scaling law; Crossover Behavior; Polynomial Trend; Constant Shift; Linear Trend.

\end{abstract}


\maketitle

\end{CJK*}

\section{Introduction}
\label{s1:Introduction}

Many natural, social and technological systems exhibit complex behavior characterized by long-term power-law correlations \cite{Sornette-2004}. There are a wealth of methods developed to determine the correlation strength in long-term correlated time series \cite{Taqqu-Teverovsky-Willinger-1995-Fractals,Montanari-Taqqu-Teverovsky-1999-MCM,Audit-Bacry-Muzy-Arneodo-2002-IEEEtit,Delignieres-Ramdani-Lemoine-Torre-Fortes-Ninot-2006-JMPsy,Kantelhardt-2009-ECSS}. The most classic method is Hurst analysis or rescaled range analysis (R/S) \cite{Hurst-1951-TASCE,Mandelbrot-Wallis-1969b-WRR}. Other popular methods include wavelet transform module maxima (WTMM) approaches \cite{Holschneider-1988-JSP,Muzy-Bacry-Arneodo-1991-PRL,Bacry-Muzy-Arneodo-1993-JSP,Muzy-Bacry-Arneodo-1993-PRE,Muzy-Bacry-Arneodo-1994-IJBC,Audit-Bacry-Muzy-Arneodo-2002-IEEEtit}, detrended fluctuation analysis (DFA) \cite{Peng-Buldyrev-Havlin-Simons-Stanley-Goldberger-1994-PRE} based on the fluctuation analysis (FA) \cite{Peng-Buldyrev-Goldberger-Havlin-Sciortino-Simons-Stanley-1992-Nature}, detrending moving average analysis (DMA) \cite{Alessio-Carbone-Castelli-Frappietro-2002-EPJB,Arianos-Carbone-2007-PA,Carbone-2009-IEEE} based on the moving average or mobile average technique \cite{Vandewalle-Ausloos-1998-PRE}, and so on. These methods have been generalized in many directions, such as objects in high dimensions \cite{Gu-Zhou-2006-PRE,Carbone-2007-PRE,AlvarezRamirez-Echeverria-Rodriguez-2008-PA,Turk-Carbone-Chiaia-2010-PRE}, detrended cross-correlation analysis and its variants for two time analysis \cite{Jun-Oh-Kim-2006-PRE,Podobnik-Stanley-2008-PRL,Zhou-2008-PRE,Podobnik-Horvatic-Petersen-Stanley-2009-PNAS,Horvatic-Stanley-Podobnik-2011-EPL,Jiang-Zhou-2011-PRE,Kristoufek-2011-EPL}, detrended partial cross-correlation analysis for multivariate time series \cite{Liu-2014,Yuan-Fu-Zhang-Piao-Xoplaki-Luterbacher-2015-SR,Qian-Liu-Jiang-Podobnik-Zhou-Stanley-2015-PRE}, multifractal analysis \cite{Kantelhardt-Zschiegner-KoscielnyBunde-Havlin-Bunde-Stanley-2002-PA,Gu-Zhou-2010-PRE}, and so on.

An important issue is to compare the performance and relative merits of these estimators, which has been conducted through extensive numerical experiments. With time series generated from the modified Fourier filtering method \cite{Makse-Havlin-Schwartz-Stanley-1996-PRE}, Xu et al. found that DFA is superior to different DMA variants \cite{Xu-Ivanov-Hu-Chen-Carbone-Stanley-2005-PRE}. Bashan et al. observed that CDMA performs comparably well as DFA for long time series with weak trends and slightly outperforms DFA for short data with weak trends \cite{Bashan-Bartsch-Kantelhardt-Havlin-2008-PA}. Based on fractional Gaussian noises (FGNs) generated from the Davies-Harte algorithm \cite{Davis-Harte-1987-Bm} and fractional Brownian motions by summing the FGNs, Serinaldi found that DFA and DMA have comparable performances \cite{Serinaldi-2010-PA}. Jiang and Zhou reported that DFA and CDMA perform similarly and both of them outperform the BDMA and FDMA methods \cite{Jiang-Zhou-2011-PRE}, in which the FBMs are generated using the Fourier-based Wood-Chan algorithm \cite{Wood-Chan-1994-JCGS}. Huang et al. reported comparative performances of FA and DFA for FBMs with $H = 1/3$  \cite{Huang-Schmitt-Hermand-Gagne-Lu-Liu-2011-PRE}, which were generated with the Wood-Chan algorithm \cite{Wood-Chan-1994-JCGS}. Bryce and Sprague reported that FA outperforms DFA, for FGNs with $H=0.3$ \cite{Bryce-Sprague-2012-SR}, which were generated using the Davies-Harte algorithm \cite{Davis-Harte-1987-Bm}, while Shao et al. found that CDMA has the best
performance, DFA is only slightly worse in some situations, and FA performs the worst \cite{Shao-Gu-Jiang-Zhou-Sornette-2012-SR}. It is not unreasonable that the conclusions are mixed because different studies used different time series generators and different lengths.

Time series in real complex systems usually contains various forms of trends and nonstationarity. Hence, another important issue concerns the effects of trends and nonstationarity on the scaling behaviors of different methods. Montanari et al. investigated the effects of periodicity on several methods such as aggregated variance method, Higuchi's method, R/S analysis, periodogram method, Whittle method, and so on \cite{Montanari-Taqqu-Teverovsky-1999-MCM}. Kantelhardt et al. studied the effects of polynomial trends and oscillatory trends on the different orders of DFA \cite{Kantelhardt-KoscielnyBunde-Rego-Havlin-Bunde-2001-PA}. Hu et al systematically studied the effects of linear, periodic, and power-law trends on DFA \cite{Hu-Ivanov-Chen-Carpena-Stanley-2001-PRE}. Chen et al. considered the presence of non-stationarity and nonlinear filters in the DFA analysis \cite{Chen-Ivanov-Hu-Stanley-2002-PRE,Chen-Hu-Carpena-BernaolaGalvan-Stanley-Ivanov-2005-PRE}. Ma et al. researched the effect of missing extreme data on DFA \cite{Ma-Bartsch-BernaolaGalvan-Yoneyama-Ivanov-2010-PRE}. Song and Shang investigated the effects of five trends based on linear and nonlinear filters on multifractal DCCA based on DFA \cite{Song-Shang-2011-Fractals}. In most cases, a crossover appears in the scaling behavior of the DFA fluctuation functions, which makes it difficult to estimate the intrinsic long-term correlations in time series. Many efforts have been made to reduce or minimize these effects on the DFA method \cite{Nagarajan-Kavasseri-2005-CSF,Nagarajan-Kavasseri-2005-IJBC,Nagarajan-Kavasseri-2005-PA,Xu-Shang-Kamae-2009-CSF,Shang-Lin-Liu-2009-PA,Gao-Hu-Tung-2011-PLoS1,Lin-Shang-2011-Fractals,Zhao-Shang-Zhao-Wang-Tao-2012-CSF}.

However, studies on the effect of trends on the detrending moving average analysis are rare, although DMA is ``The Method of Choice'' as DFA \cite{Shao-Gu-Jiang-Zhou-Sornette-2012-SR}. To our knowledge, one such study is to minimize the effect of period trends on the DMA method \cite{Lin-Shang-2011-Fractals}. In this work, we aim at contributing this direction by investigating the effects of polynomial trends on the scaling behavior of DMA methods. We derive analytically the results for constant shift and linear trend and confirm these results using numerical experiments.

\section{Detrending moving average algorithms}
\label{s2:Algorithms}

The algorithms of the detrending moving average analysis are described as follows \cite{Alessio-Carbone-Castelli-Frappietro-2002-EPJB,Carbone-Castelli-2003-SPIE,Carbone-Castelli-Stanley-2004-PA,Carbone-Stanley-2004-PA,Carbone-Castelli-Stanley-2004-PRE,Xu-Ivanov-Hu-Chen-Carbone-Stanley-2005-PRE,Arianos-Carbone-2007-PA,Carbone-2009-IEEE}.

{\em{Step 1}}. Consider a time series $x(t)$, $t=1,2,\cdots,N$. We construct the sequence of cumulative sums
\begin{equation}
  X(t)=\sum_{i=1}^{t}{x(i)}, ~~t=1, 2, \cdots, N.
  \label{Eq:DMA:1D:X}
\end{equation}

{\em{Step 2}}. Consider a box $[t-s_1,t+s_2]$ of size $s=s_1+s_2+1$, where $s_1=\lceil(s-1)(1-\theta)\rceil$, $s_2=\lfloor(s-1)\theta\rfloor$, $\lfloor{x}\rfloor$ is the largest integer smaller than $x$, $\lceil{x}\rceil$ is the smallest integer larger than $x$, and $\theta$ is the position parameter with the value varying in the range $[0,1]$. Calculate the moving average function $\widetilde{X}(t)$ in a moving window \cite{Arianos-Carbone-2007-PA},
\begin{equation}
  \widetilde{X}(t)=\frac{1}{s}\sum_{k=-s_2}^{s_1}X(t-k).
  \label{Eq:DMA:1D:X:MA}
\end{equation}
Hence, the moving average function considers $s_1$ data points in the past and $s_2$ points in the future. We consider three special cases in this paper. The first case $\theta=0$ refers to the backward moving average \cite{Xu-Ivanov-Hu-Chen-Carbone-Stanley-2005-PRE}, in which the moving average function $\widetilde{X}(t)$ is calculated over all the past $n-1$ data points of the signal. The second case $\theta=0.5$ corresponds to the centered moving average \cite{Xu-Ivanov-Hu-Chen-Carbone-Stanley-2005-PRE}, where $\widetilde{X}(t)$ contains half past and half future information in each window. The third case $\theta=1$ is called the forward moving average, where $\widetilde{X}(t)$ considers the trend of $n-1$ data points in the future.

{\em{Step 3}}. Detrend the signal series by removing the moving average function $\widetilde{X}(i)$ from $X(i)$, and obtain the residual sequence $\epsilon(i)$ through
\begin{equation}
  \epsilon(i)=X(i)-\widetilde{X}(i),
  \label{Eq:DMA:1D:epsilon}
\end{equation}
where $n-\lfloor(s-1)\theta\rfloor\leqslant{i}\leqslant{N-\lfloor(s-1)\theta\rfloor}$.

{\em{Step 4}}. The residual series $\epsilon(i)$ is divided into $N_s$ disjoint segments with the same size $s$, where $N_s=\lfloor{N}/n-1\rfloor$. Each segment can be denoted by $\epsilon_v$ such that $\epsilon_v(i)=\epsilon(l+i)$ for $1\leqslant{i}\leqslant{s}$, where $l=(v-1)s$. The root-mean-square function $F_v(s)$ with the window size $s$ can be calculated by
\begin{equation}
  F_v^2(s)=\frac{1}{s}\sum_{i=1}^{s}\epsilon_v^2(i).
  \label{Eq:DMA:1D:Fv2}
\end{equation}

{\em{Step 5}}. Varying the values of segment size $s$, we can determine the power-law relation between the function $F(s)$ and the size scale $s$, which reads
\begin{equation}
  F(s)\sim bs^{H}.
  \label{Eq:dma_m_h}
\end{equation}

\section{Polynomial trends}

Consider a signal composed of a signal $x(t)$ with zero mean and an additive trend $u(t)$
\begin{equation}
  \label{Eq:z_def}
  z(t)=x(t)+u(t)
\end{equation}
The profile of $z(t)$ is the sum of the profiles of $x(t)$ and $u(t)$:
\begin{equation}
  \label{Eq:Z_def}
  Z(t)=X(t)+U(t)
\end{equation}
and the moving average at time $t$ is
\begin{equation}
\label{Eq:Z_ma_def}
\widetilde{Z}(t)=\widetilde{X}(t)+\widetilde{U}(t)
\end{equation}
When $q=2$, the overall fluctuation is
\begin{equation}
\begin{aligned}
\label{Eq:Fz(s)_def}
F^{2}_{z}(s)&=\sum_{i=1}^{N} [Z(t)-\tilde{Z}(t)]^{2} \\
&=\sum_{i=1}^{N} [X(t)-\tilde{X}(t)+U(t)-\tilde{U}(t)]^{2} \\
&=F^{2}_{x}(s)+F^{2}_{u}(s)+2\sum_{i=1}^{N} [\epsilon_{x}(t)\epsilon_{u}(t)]
\end{aligned}
\end{equation}
where $\epsilon_{x}(t)=X(t)-\tilde{X}(t)$ and $\epsilon_{u}(t)=U(t)-\tilde{U}(t)$. If $\epsilon_{x}(t)$ and $\epsilon_{u}(t)$ are uncorrelated, we have
\begin{equation}
\label{Eq:Fz(s)_superposition}
F_{z}^{2}(s)=F_{x}^{2}(s)+F_{u}^{2}(s)
\end{equation}
which is the superposition rule \cite{Hu-Ivanov-Chen-Carpena-Stanley-2001-PRE}.

We consider polynomial trends added to the increments series:
\begin{equation}
\label{Eq:polyu_def}
u(t)=\sum_{p=0}^{m} a_{p} t^{p}
\end{equation}
and the cumulative sum is
\begin{equation}
\label{Eq:polyU_def}
U(t)=\sum_{i=1}^{t} u(i)=\sum_{p=0}^{m} a_{p} \sum_{i=1}^{t} t^{p}
\end{equation}

According to Faulhaber's formula, the sum of powers $\sum_{i=1}^{t} i^{p}$ can be expressed as follows:
\begin{equation}
\label{Eq:sum_i_Ff}
\sum_{i=1}^{t} i^{p}=\frac{1}{p+1} \sum_{k=0}^{p} C_{p+1}^{k} B_{k} t^{p+1-k}
\end{equation}
where the coefficients $B_{k}$ are the Bernoulli numbers. For $p$=0, 1, 2 and 3, we have $\sum_{i=1}^{t}i^{0}=t$, $\sum_{i=1}^{t}i^{1}={(t^2+t)/2}$, $\sum_{i=1}^{t}i^{2}={(2t^3+3t^2+t)/6}$ and
$\sum_{i=1}^{t}i^{3}={(t^4+2t^3+t^2)/4}$. It follows that
\begin{equation}
\begin{split}
  U(t)&=\left(a_0+\frac{1}{2}a_1+\frac{1}{6}a_{2}\right)t + \left(\frac{1}{2}a_1+\frac{1}{2}a_{2}+\frac{1}{4}a_{3}\right)t^{2}\\
  &+\left(\frac{1}{3}a_2+\frac{1}{2}a_3\right)t^{3}+\frac{1}{4}a_{3}t^{4}
\label{Eq:U_Ff}
\end{split}
\end{equation}

The moving average at $t$ from $t-s_{1}$ to $t+s_{2}$ is
\begin{equation}
  \tilde {Z}(t)=\frac{1}{s} \sum_{k=-s_{2}}^{s_{1}} Z(t-k)= \frac{1}{s} \sum_{k=t-s_{1}}^{t+s_{2}} Z(k)
  \label{Eq:Z_MA}
\end{equation}
where $s=s_{1}+s_{2}-1$ is the window size.

\section{Constant shift: The case of $p=0$}
\label{S1:p=0}

\subsection{Analytical results}
\label{S2:p=0:Analytic}

In this case, we consider $a_1=a_{2}=a_{3}=0$. The trend is a constant shift
\begin{equation}
  u(t)=a_0
  \label{Eq:u p0}
\end{equation}
The cumulative sum, or the profile, is
\begin{equation}
  U(t)=\sum_{i=1}^{t} a_0t
  \label{Eq:U_p0}
\end{equation}
The moving average at $t$ obtained from $t-s_{1}$ to $t+s_{2}$ for window size $s$ is
\begin{equation}
  \widetilde {U}(t)=a_0\left(t+\frac{s-1}{2}-s_{1}\right)
  \label{Eq:U_ma_p0}
\end{equation}
where $s_{1}+s_{2}+1=s$. Since $s_{1}=(s-1)(1-\theta)$ when $\theta=0,0.5$ and 1 (note that $s$ should be odd), we have
\begin{equation}
  \widetilde {U}(t)=a_0\left[t+\frac{(2\theta-1)(s-1)}{2}\right]
  \label{Eq:U_MA_p0}
\end{equation}
and the residual of the trend after removing the moving average is
\begin{equation}
  \epsilon _{u}(t)=a_0 \frac{(2\theta-1)(s-1)}{2}
  \label{Eq:epsilon_p0}
\end{equation}
which is a constant for a given window size $s$. Hence the superposition rule holds.

When $\theta=0.5$, we have
\begin{equation}
  F_{z}^{2} (s)=\frac{1}{N} \sum_{i=1}^{N} \epsilon^{2}(t)= F_{x}^{2} (t),
  \label{Eq:Fz:p=0:CDMA}
\end{equation}
which is independent of the constant shift term $a_0$. It indicates that, if $x$ is a fractional Gaussian noise, there is no crossover in the $F_{z}$ scaling.

When $\theta=0$ and $\theta=1$, the detrended fluctuation is
\begin{equation}
  F_{z}^{2} (s)=F_{x}^{2} (t)+ \frac{a_0^{2}(s-1)^{2}}{4},
  \label{Eq:Fz:p=0:BDMA:FDMA}
\end{equation}
which depends on $a_0$ and $s$. There is a crossover $s=s_{\times}$ in $F_{z}(s)$. For $s<s_{\times}$, the behavior of $F_{z}(s)$ is very close to the behavior of $F_{x}(s)$, while for $s>s_{\times}$, the behavior of $F_{z}(s)$ is very close to the behavior of $F_{u}(s)$. The crossover scale $s_{\times}$ is the solution to the following equation
\begin{equation}
\label{Eq:FxvsFu_superposition}
F_{x} (s)= F_{u} (s).
\end{equation}
It follows that $bs^H=a_0(s-1)/2$. When $s\gg1$, we have
\begin{equation}
  \label{Eq:sx:p0_BDMA&FDMA}
  s_{\times}=\left(\frac{2b}{a_0} \right)^{1/(1-H)},
\end{equation}
which shows that $s_{\times}$ is a power-law function of $a_0$ with the exponent being $-1/(1-H)$.

\subsection{Numerical experiments}
\label{S2:p=0:Numerical}

\begin{figure*}[t]
  \centering
  \includegraphics[width=0.32\textwidth,height=0.25\textwidth]{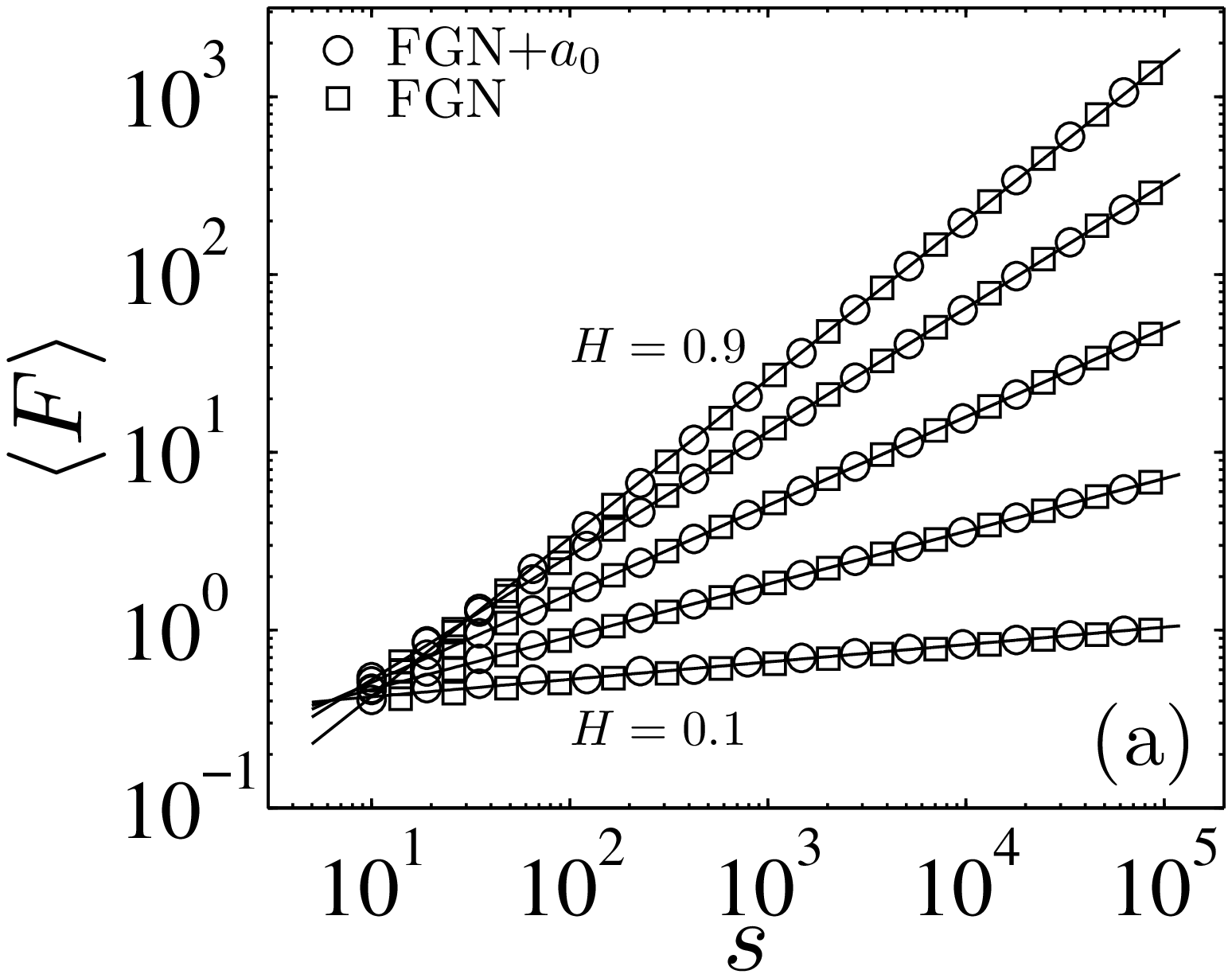}
  \includegraphics[width=0.32\textwidth,height=0.25\textwidth]{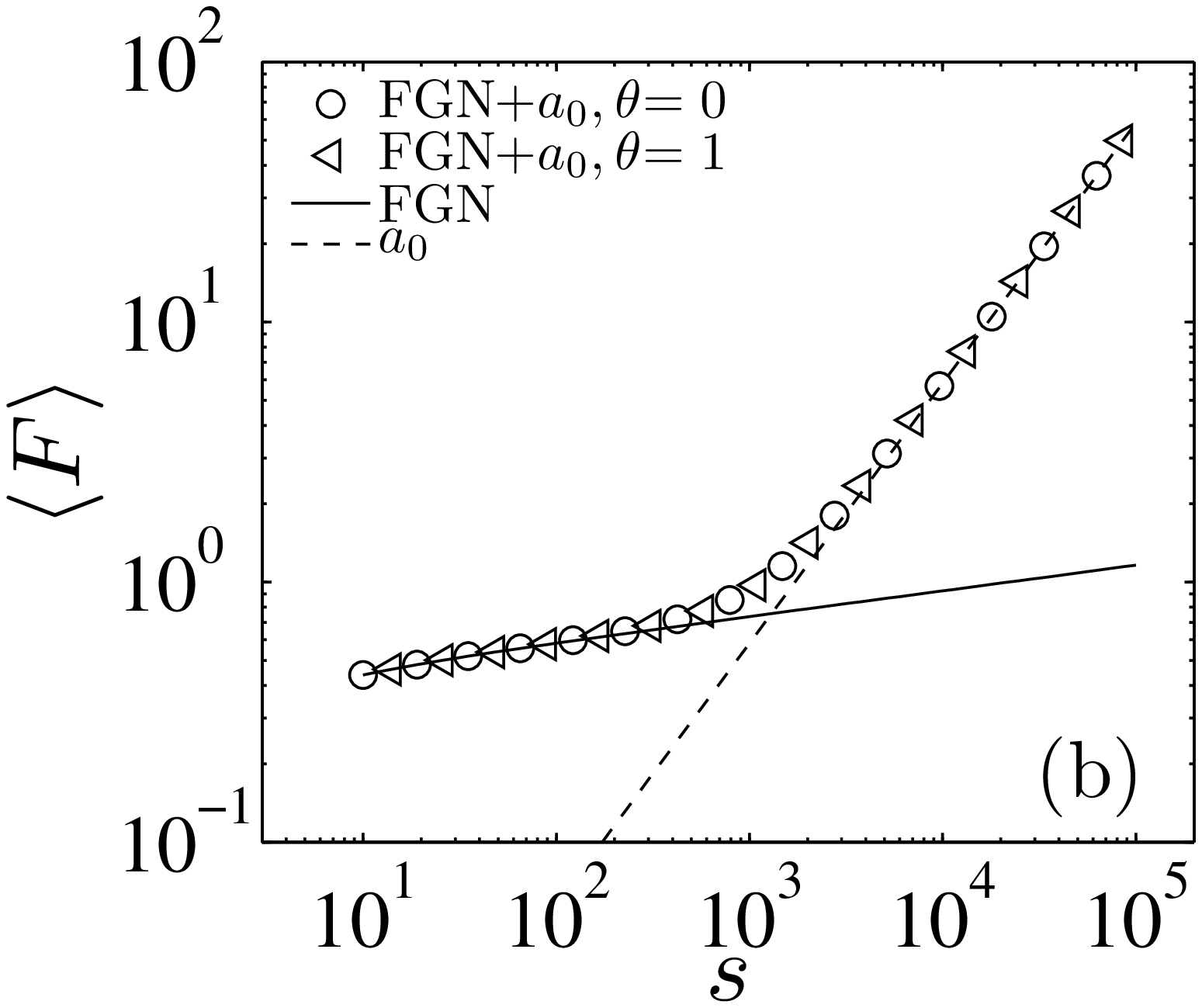}
  \includegraphics[width=0.32\textwidth,height=0.25\textwidth]{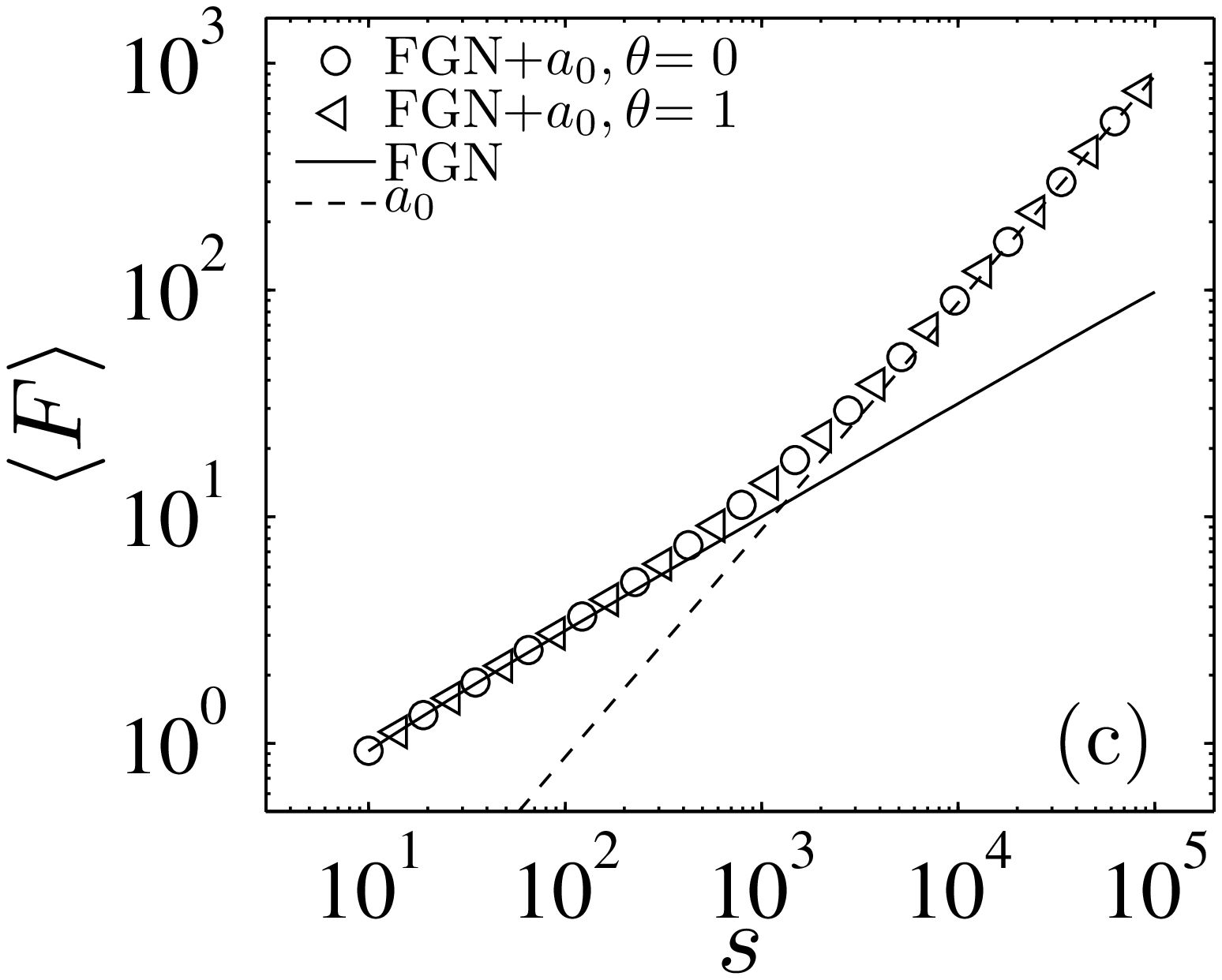}\\
  \includegraphics[width=0.32\textwidth,height=0.25\textwidth]{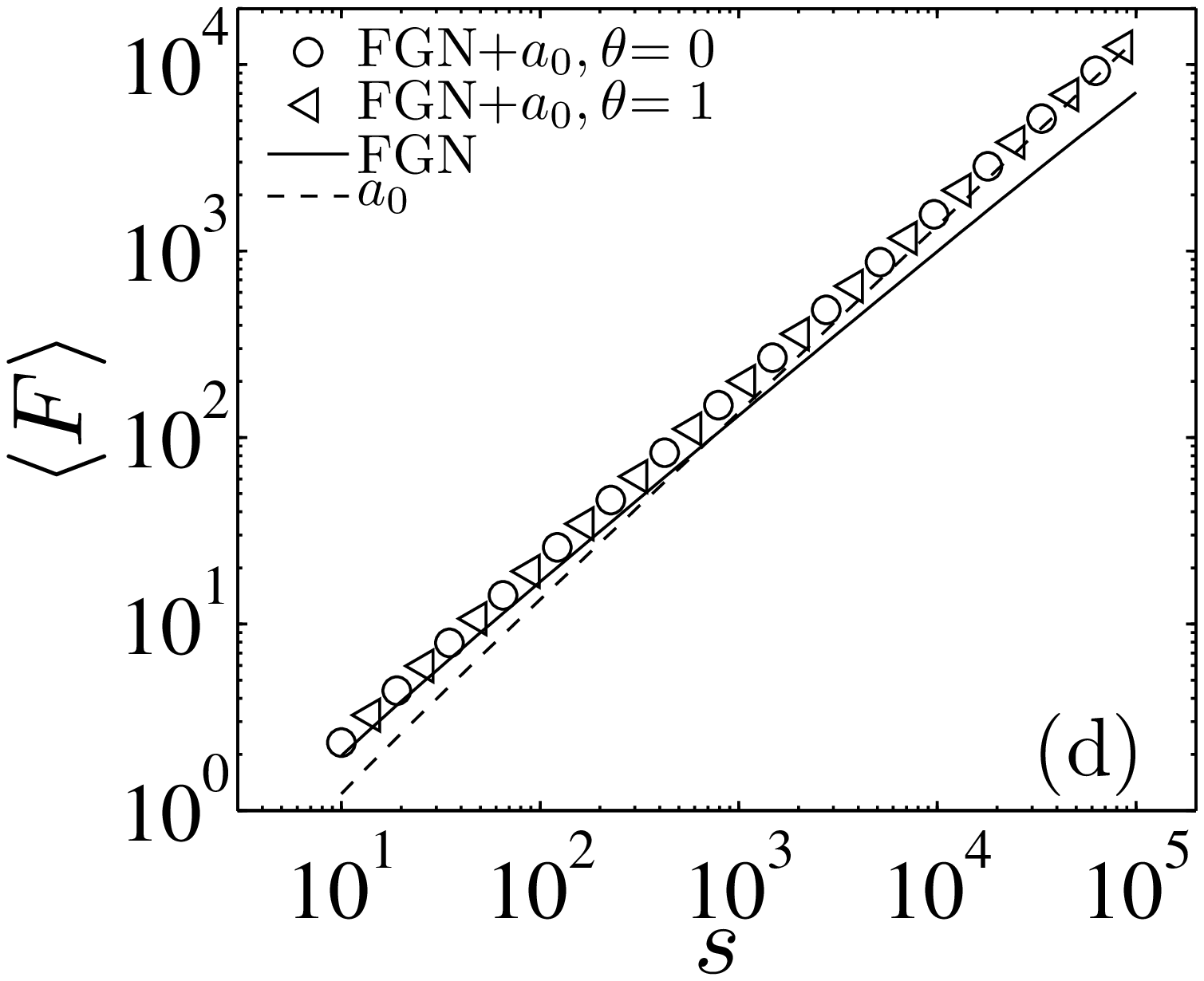}
  \includegraphics[width=0.32\textwidth,height=0.25\textwidth]{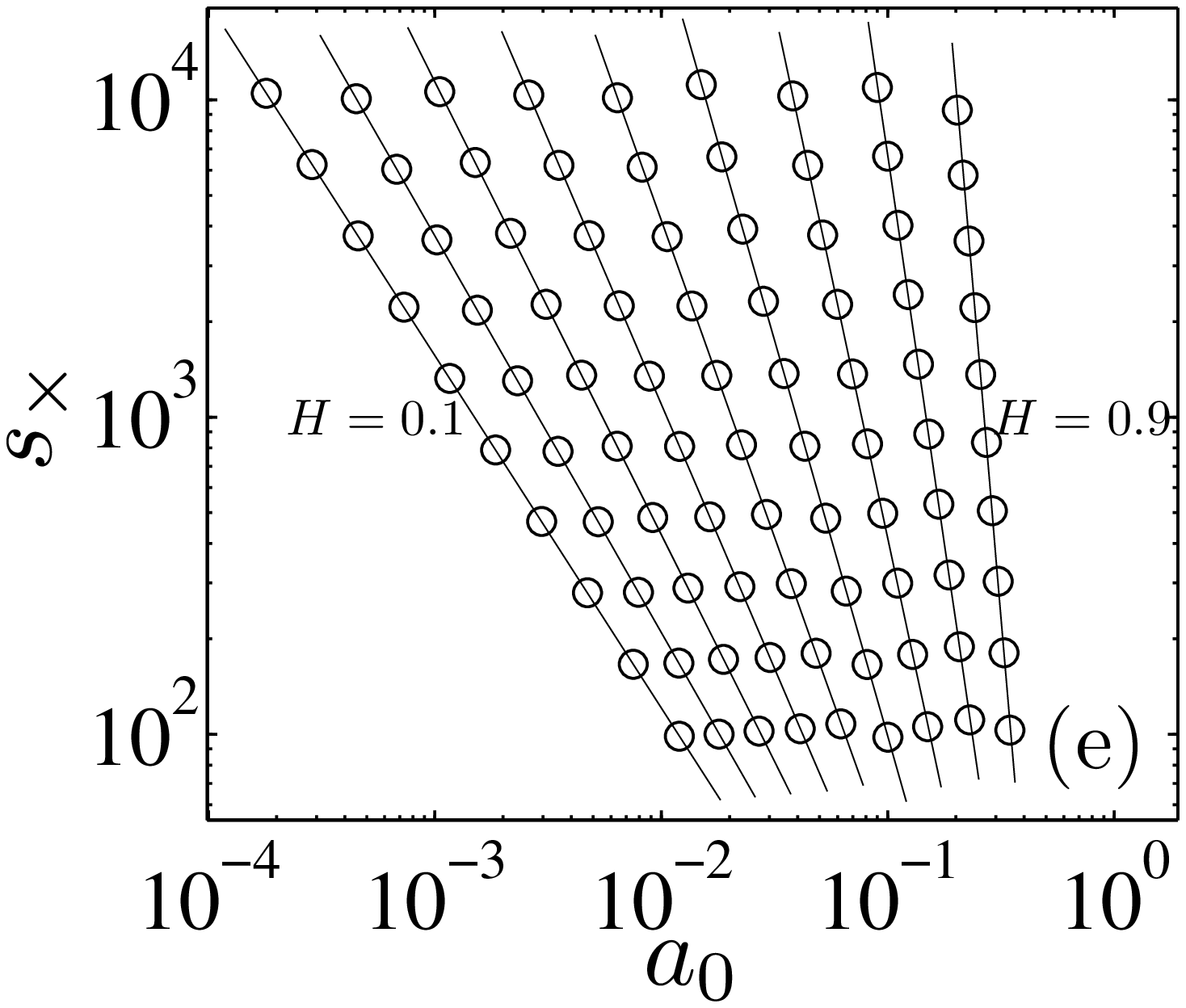}
  \includegraphics[width=0.32\textwidth,height=0.25\textwidth]{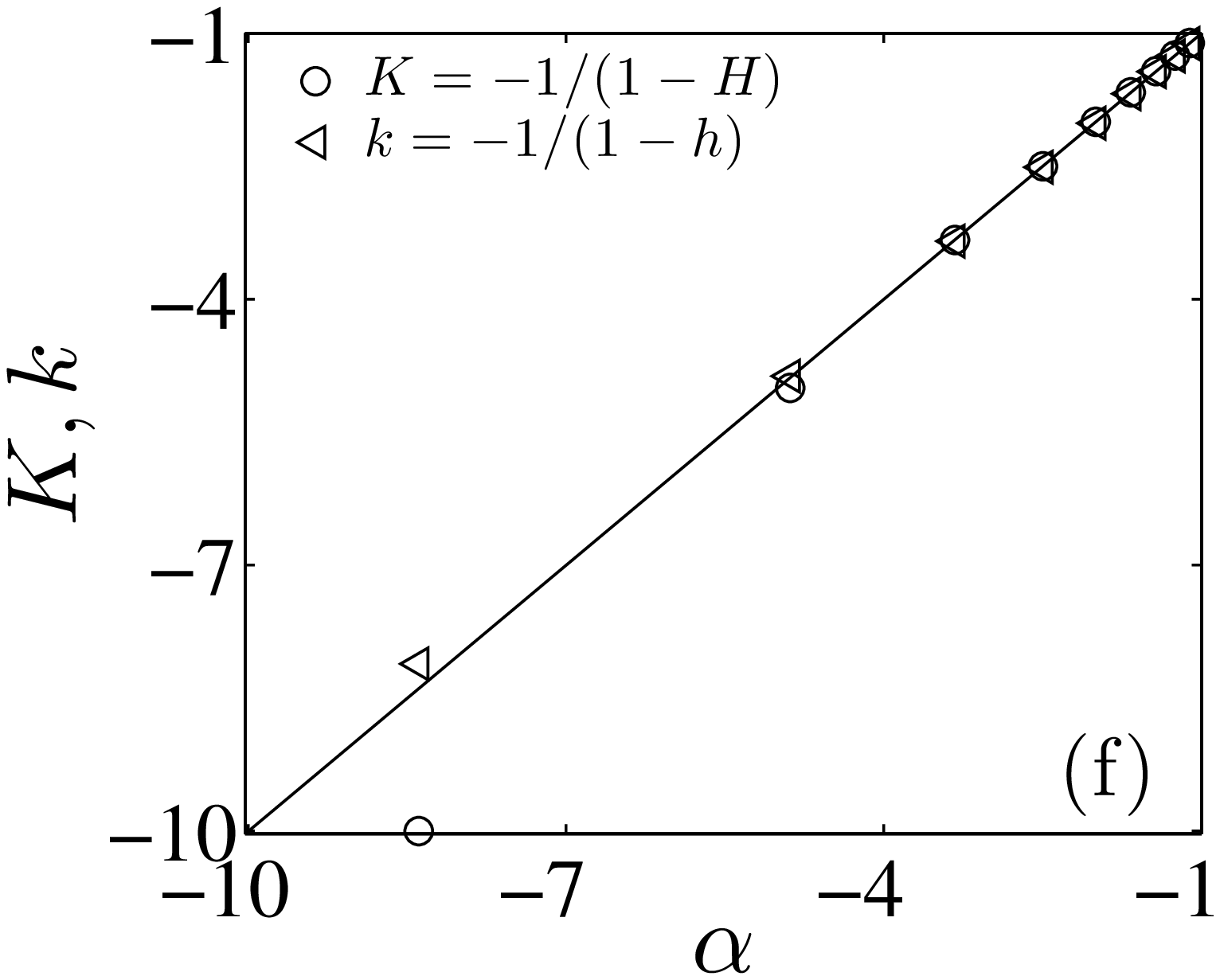}
  \caption{Effect of constant shift on the DMA algorithms. Each curve in (a-d) represents a fluctuation function averaged over 50 repeated simulations. (a) Log-log plots of $\langle F \rangle$ of the FGNs with different Hurst indexes and the FGNs with constant shift $a_0=0.2$ in the increments against $s$ for the CDMA method. The Hurst index $H$ varies from 0.1 (bottom) to 0.9 (top) with a step of 0.1. (b) Log-log plots of $\langle F \rangle$ against $s$ using the BDMA and FDMA methods for $H_{\rm{in}}=0.1$ and $a_0=0.0012$. (c) Log-log plots of $\langle F \rangle$ against $s$ using the BDMA and FDMA methods for $H=0.5$ and $a_0=0.0176$. (d) Log-log plots of $\langle F \rangle$ against $s$ using the BDMA and FDMA methods for $H=0.9$ and $a_0=0.257$. (e) Power-law dependence of the crossover scale $s_{\times}$ on the constant shift $a_0$ using the BDMA method for different Hurst indexes, varying from 0.1 (left) to 0.9 (right) with a step of 0.1. (f) Validation of Eq.~(\ref{Eq:sx:p0_BDMA&FDMA}). The crossover exponent $\alpha$ is the power-law exponent in (e) and $K=-1/(1-H)$ and $k=-1/(1-h)$, where $H$ is the input Hurst index for the generation of FGNs and $h$ is the estimated Hurst index of the generated FGNs using BDMA.}
  \label{Fig:DMA:ConstantShift}
\end{figure*}

We perform numerical simulations to verify the correctness of the main results, Eq.~(\ref{Eq:Fz:p=0:CDMA}) and Eq.~(\ref{Eq:sx:p0_BDMA&FDMA}), derived in Sec.~\ref{S2:p=0:Analytic}.
We employ the Davies-Harte algorithm \cite{Davis-Harte-1987-Bm} to generate fractional Gaussian noise (FGN) with given Hurst indexes $H$. There are other comparable generators such as the wavelet-based FMB generator \cite{Abry-Sellan-1996-ACHA} and the random midpoint displacement algorithm \cite{Mandelbrot-1983}. However, the Davies-Harte algorithm performs slightly better \cite{Shao-Gu-Jiang-Zhou-Sornette-2012-SR}. In our simulations, we consider different Hurst indexes $H$, which range from 0.1 to 0.9 with a step of 0.1. For each $H$, we generate 50 FGN time series $x(t)$ of length $10^6$. The constant shift $a_0$ is added to each point of the each FGN series. The DMA fluctuation functions presented below are averaged over the 50 realizations.

Figure \ref{Fig:DMA:ConstantShift}a illustrates the averaged fluctuation functions obtained from the CMDA method for different $H$ values. The two curves for the original FGN time series and for the shifted FGNs with $a_0=0.2$ overlap excellently. In addition, all the curves have excellent power-law forms with the slopes being the corresponding $H$ values. Changing the value of $a_0$ has no impact on the results. Therefore, Fig.~\ref{Fig:DMA:ConstantShift}a verifies Eq.~(\ref{Eq:Fz:p=0:CDMA}) exactly.

Figure \ref{Fig:DMA:ConstantShift}b shows the average fluctuation functions of the constantly shifted FGNs with $a_0=0.0012$ and $H=0.1$ obtained from the BDMA method and the FDMA method. The two curves overlap nicely. The fluctuation functions exhibit a clear crossover. When $s \ll s_{\times}$, the fluctuation functions overlap with the fluctuation function of FGNs with the slope being $H=0.1$. When $s\gg s_{\times}$, the fluctuation functions overlap with the fluctuation function of the constant $a_0$ with the slope being $H=1$. These observations are consistent with Eq.~(\ref{Eq:Fz:p=0:BDMA:FDMA}). Figure \ref{Fig:DMA:ConstantShift}c and Fig.~\ref{Fig:DMA:ConstantShift}d show the results for $H=0.5$ and $a_0=0.0176$ and for $H=0.9$ and $a_0=0.257$, respectively. These results are also consistent with the prediction of Eq.~(\ref{Eq:Fz:p=0:BDMA:FDMA}).

%

We can determine the crossover scale $s_{\times}$ by two methods. The first one is to use $F_x(s_{\times})=F_u(s_{\times})$, determining the intersection point $(s_{\times},F_x(s_{\times}))$ of the solid line and the dashed line in each plot (Fig.~\ref{Fig:DMA:ConstantShift}b-Fig.~\ref{Fig:DMA:ConstantShift}d). However, this method uses a priori information about the underlying FGNs and the constant shift. An alternative method is as described below. We pinpoint the point $s_m$ on the fluctuation curve that is the farthest from the line connect the two endpoints. We perform a linear fit the first few point from the right point to a point that is in the middle of the right point and $(s_m,F_z(s_m))$ to obtain a first straight line and similarly a second straight line based on the right part of the $F_z$ curve. The crossover scale is determined by the intersection of these two straight lines. In this procedure, the choices of the right point $s_m$ for the left part of the fluctuation function and the left point of the right part of the fluctuation function can vary, which does not influence the determination of $s_{\times}$. In our analysis, we simply use the five left-most data points and the five right-most data points in the linear regressions and obtain the intersection of the two regressed lines treating as $s_\times$. Note that there are 60 points in each fluctuation function. Figure \ref{Fig:DMA:ConstantShift}e shows the dependence of $s_{\times}$ as a function of $a_0$ for different $H$ values. For every $H$ value, we observe a nice power-law relationship:
\begin{equation}
  s_{\times} \sim a_0^{\alpha},
  \label{Eq:p0:sx:a0:alpha}
\end{equation}
which is consistent with the power-law form expressed in Eq.~(\ref{Eq:sx:p0_BDMA&FDMA}).

Figure \ref{Fig:DMA:ConstantShift}f shows that the lines become steeper for larger Hurst indexes. We fit the data points for each $H$ to estimate the power-law exponent $\alpha$. We then define and calculate the following two quantities:
\begin{equation}
  K=-1/(1-H)
\end{equation}
and
\begin{equation}
  k=-1/(1-h),
\end{equation}
where $H$ is the input Hurst indexes for the synthesis of the FGNs and $h$ is the output Hurst indexes of the synthesized FGNs using the BDMA method. We plot $K$ against $\alpha$ and $k$ against $\alpha$ in Fig.~\ref{Fig:DMA:ConstantShift}f. We observe that
\begin{equation}
  K\approx k\approx\alpha,
  \label{Eq:p0:sx:K:k:alpha}
\end{equation}
except for $H=0.9$. Equations (\ref{Eq:p0:sx:a0:alpha}) and (\ref{Eq:p0:sx:K:k:alpha}) verify excellently Eq.~(\ref{Eq:sx:p0_BDMA&FDMA}).

%

The choice of $a_0$ values are not arbitrary. Due to the finite size of the generated FGNs, too large $a_0$ will result in very small $s_{\times}$ so that the resulting fluctuation function becomes a straight line with the slope being 1, while too small $a_0$ will result in very large $s_{\times}$ so that the resulting fluctuation function becomes a straight line with the slope being $H$. In both cases, the crossover cannot be recognized. In the numerical experiments, for each $H$, we use 10 $a_0$ values that are evenly spaced in the logarithmic scale. For instance, $a_0$ values are distributed in $[0.00018,0.012]$ for $H=0.1$, in $[0.0064,0.0621]$ for $H=0.5$, and in $[0.203,0.346]$ for $H=0.9$. In this way, the crossovers can be identified. The $a_0$ values used in Fig.~\ref{Fig:DMA:ConstantShift}b to Fig.~\ref{Fig:DMA:ConstantShift}d are the fifth in each of the 10 $a_0$ values.

It is clear that the numerical results illustrated in Fig.~\ref{Fig:DMA:ConstantShift} verify the analytical results in the previous subsection.

\section{Linear trend: The case of $p=1$}

\subsection{Analytical results}
\label{S2:p=1:Analytic}

We now consider the case of linear trends with $a_0=a_{2}=a_{3}=0$. The linear trend is
\begin{equation}
  \label{Eq:u_p1}
  u(t)=a_1t.
\end{equation}
The profile of $u(t)$ is
\begin{equation}
  \label{Eq:U_p1}
  U(t)=a_1(t^{2}+t)/2,
\end{equation}
and the moving average is
\begin{equation}
  \widetilde{U}(t)=a_1\left(\frac{1}{2}t^{2}+\frac{s-2s_{1}}{2} t-\frac{s s_{1}}{2}+\frac{s^{2}}{6}+\frac{s_{1}^{2}}{2}-\frac{1}{6}\right),
  \label{Eq:U_ma_p1}
\end{equation}
where $s_{1}=(s-1)(1-\theta)$. When $\theta=0,0.5,1$ (note that $s$ should be odd), we have
\begin{equation}
  \widetilde{U}(t)=a_1\left[\frac{1}{2}t^{2}+\frac{2(1-\theta)+s(2\theta-1)}{2}t\right]+L,
  \label{Eq:U_MA_p1}
\end{equation}
where
\begin{equation}
  L=a_1(s-1)(s-3s\theta+3s\theta^{2}-3\theta^{2}+6\theta-2)/6.
  \label{Eq:U_MA_p1:L}
\end{equation}
The residual series is
\begin{equation}
 \label{Eq:epsilon_p1}
 \epsilon_{u}(t)=U(t)-\widetilde{U}(t)=At-L,
\end{equation}
where
\begin{equation}
 \label{Eq:epsilon_p1:A}
 A=-a_1(s-1)(2\theta-1)/2.
\end{equation}
The detrended fluctuation is
\begin{equation}
\label{Eq:F_p1}
F_{z}^{2}(s)=\frac{1}{N} \sum_{t=1}^{N}[\epsilon_{x}(t)+\epsilon_{u}(t)]^{2}
\end{equation}
Applying the superposition rule \cite{Hu-Ivanov-Chen-Carpena-Stanley-2001-PRE} and Faulhaber's formula, we have
\begin{equation}
\begin{aligned}
\label{Eq:p=1:DMA:Fz}
F_{z}^{2}(s)&=\frac{1}{N}\sum_{t=1}^{N}[\epsilon_{x}^{2}(t)+\epsilon_{u}^{2}(t)]\\
&=F_{x}^{2}+\frac{1}{N} \sum_{t=1}^{N}(At-L)^{2}\\
&=F_{x}^{2}+L^{2}+\frac{A^2(2N^{2}+3N+1)}{6}-AL(N+1) \\
\end{aligned}
\end{equation}

When $\theta=0.5$, we have $A=0$ and $L={a_1(s^{2}-1)}/{24}$. Inserting them into Eq.~(\ref{Eq:p=1:DMA:Fz}), it follows immediately that
\begin{equation}
  \begin{aligned}
   F_{z}^{2}& =F_{x}^{2}+\left[\frac{a_1(s^{2}-1)}{24}\right]^{2}.
  \end{aligned}
  \label{Eq:p=1:CDMA:Fz}
\end{equation}
We obtain the crossover scale $s_{\times}$ as follow
\begin{equation}
  s_{\times}= \left(\frac{24b}{a_1}\right)^{{1}/{(2-H)}}.
  \label{Eq:p=1:CDMA:sx}
\end{equation}
We notice that $s_{\times}$ depends on $a_1$ but not on $N$.

When $\theta=0$, we have $L=a_1(s-1)(s-2)/6$ and $A=a_1(s-1)/2$.
The detrended fluctuation is
\begin{equation}
\begin{aligned}
  F_{z}^{2}&=F_{x}^{2}+\left[\frac{a_1(s-1)(s-2)}{6}\right]^{2}\\
         &+\frac{a_1^{2}(2N^{2}+3N+1)(s-1)^{2}}{24}\\
         &-\frac{a_1^{2}(N+1) (s-1)^{2} (s-2)}{12}
  \label{Eq:p=1:BDMA:Fz:Raw}
\end{aligned}
\end{equation}
When $1\ll s \ll N$, we have
\begin{equation}
  \begin{aligned}
    F_{z}^{2}& \approx F_{x}^{2}+a_1^{2}s^{2}\left(\frac{s^{2}}{36}-\frac{Ns}{12}+\frac{N^{2}}{12}\right)\\
    &\approx F_{x}^{2}+\frac{a_1^{2}N^{2}s^{2}}{12}
  \end{aligned}
  \label{Eq:p=1:BDMA:Fz}
\end{equation}
There is a crossover at scale
\begin{equation}
  s_{\times}= \left(\frac{\sqrt{12}b}{a_1N}\right)^{{1}/{(1-H)}},
  \label{Eq:p=1:BDMA:sx}
\end{equation}
which depends on $a_1$ and $N$.

\begin{figure*}
\centering
\includegraphics[width=0.32\textwidth]{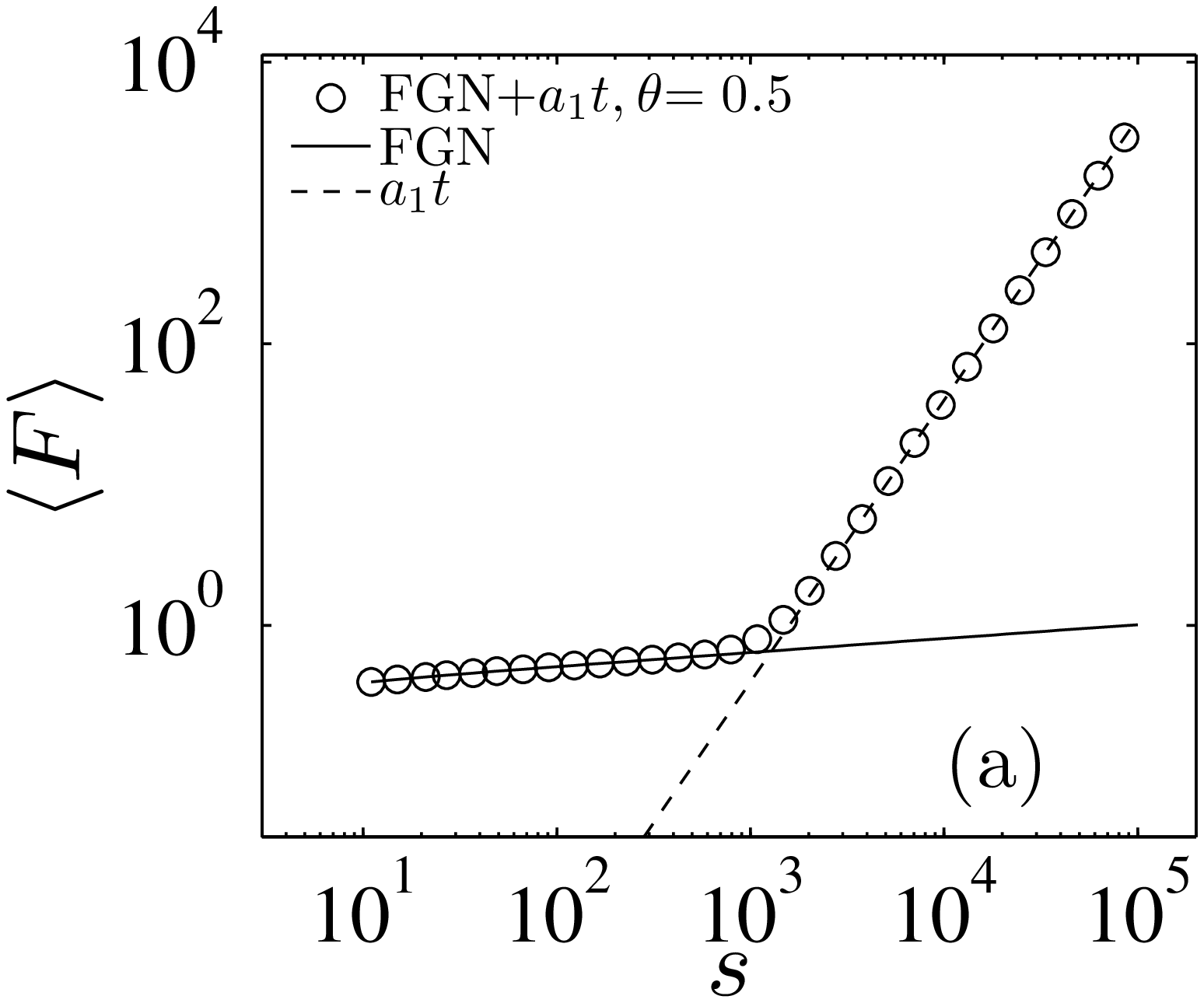}
\includegraphics[width=0.32\textwidth]{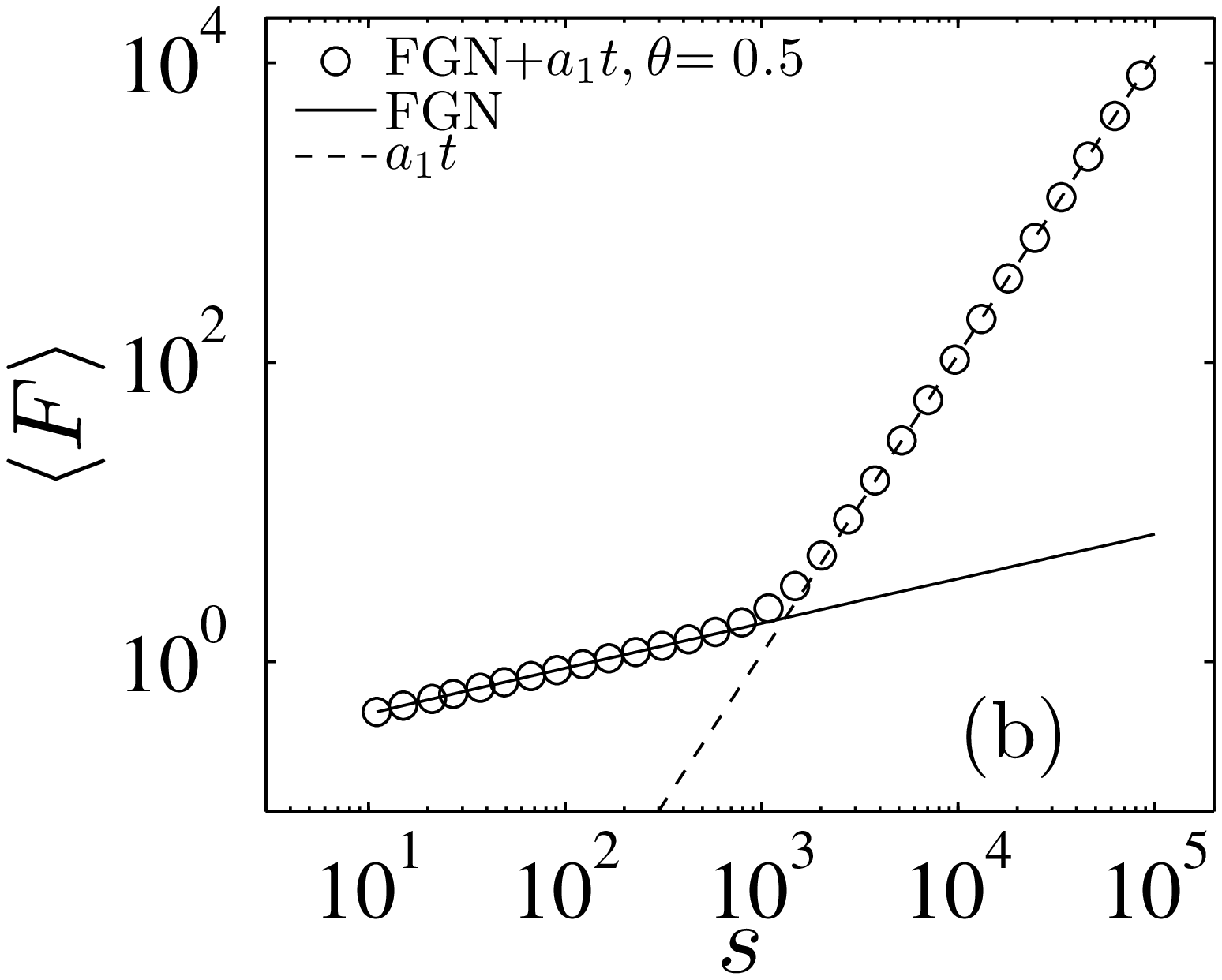}
\includegraphics[width=0.32\textwidth]{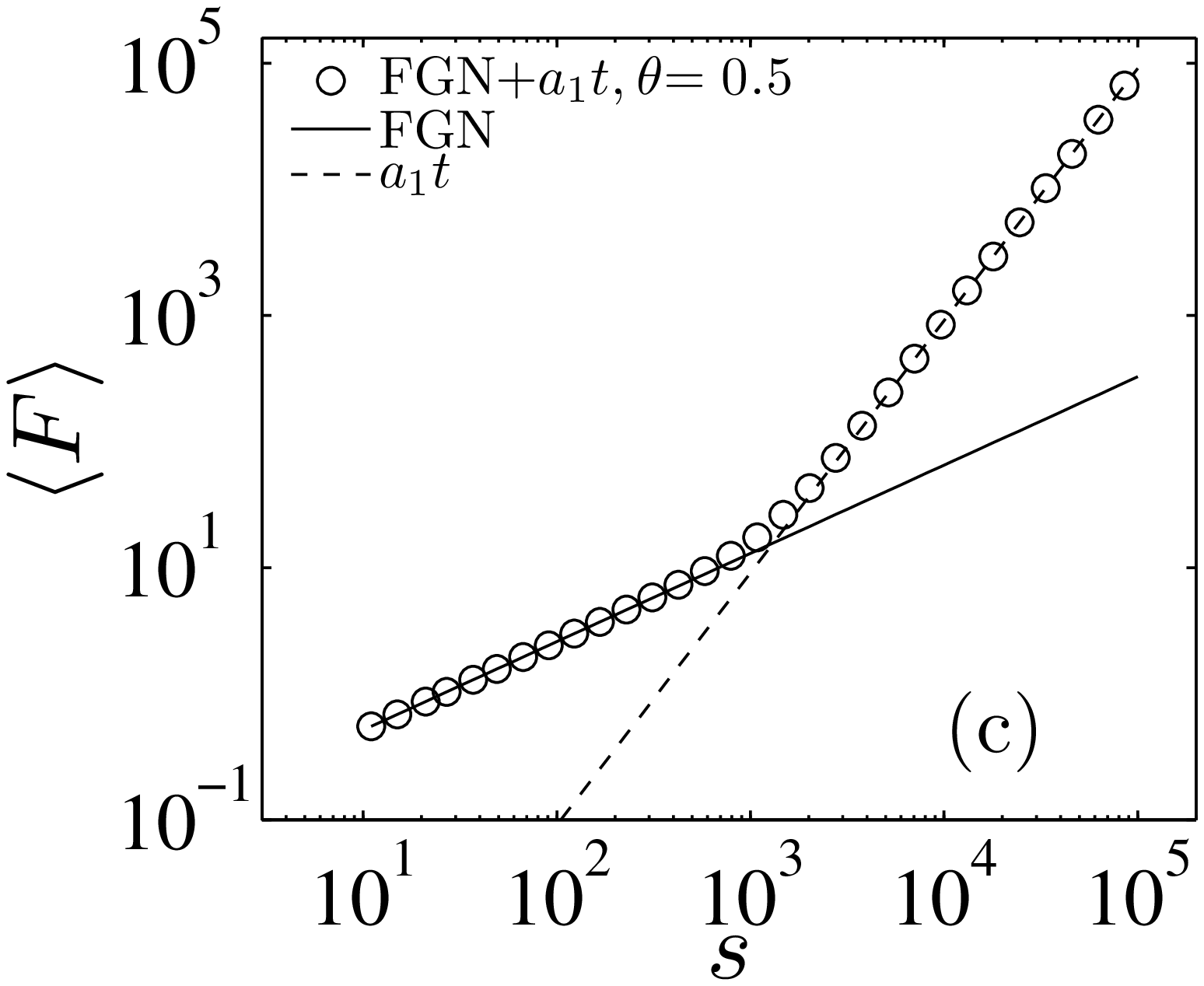}
\includegraphics[width=0.32\textwidth]{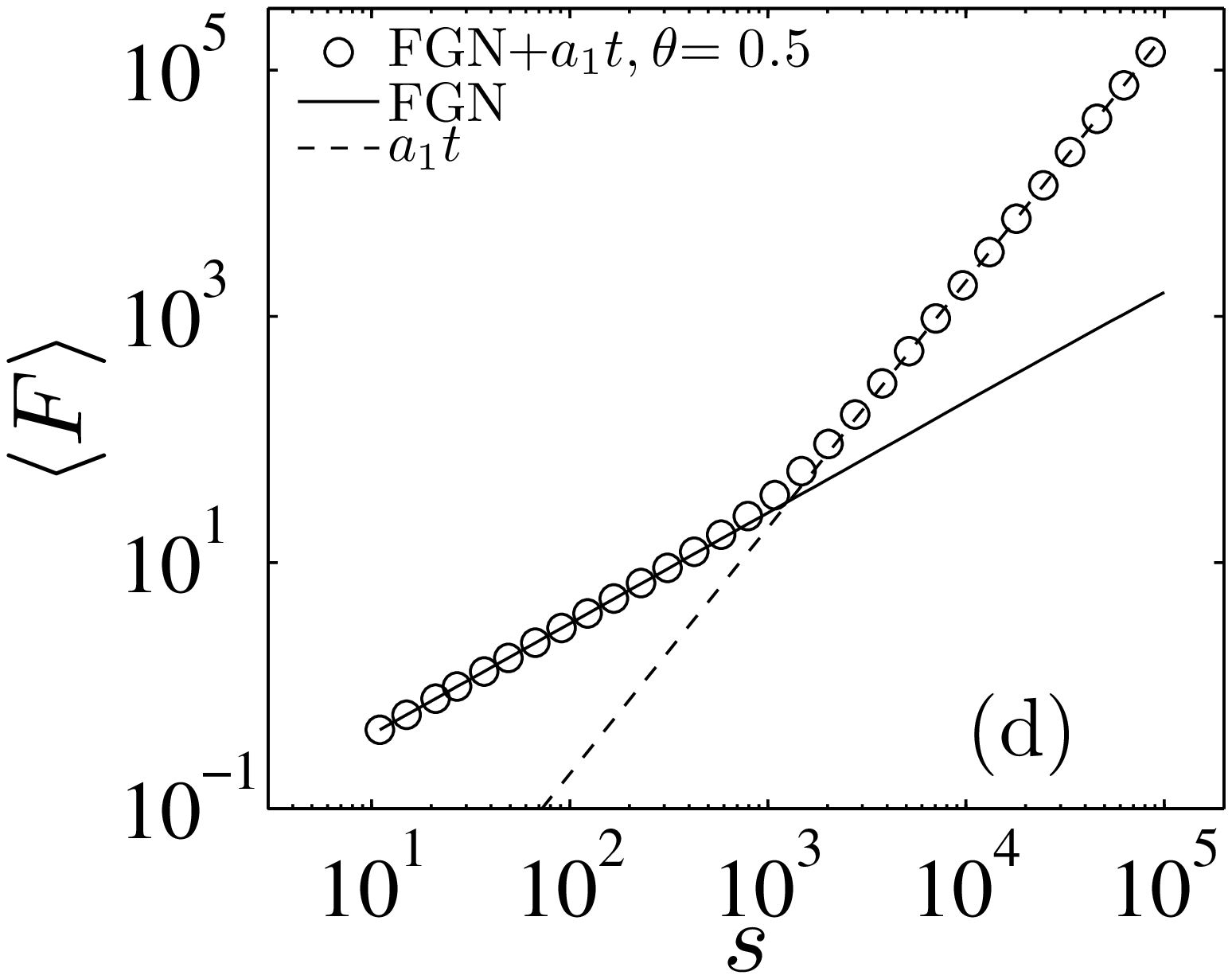}
\includegraphics[width=0.32\textwidth]{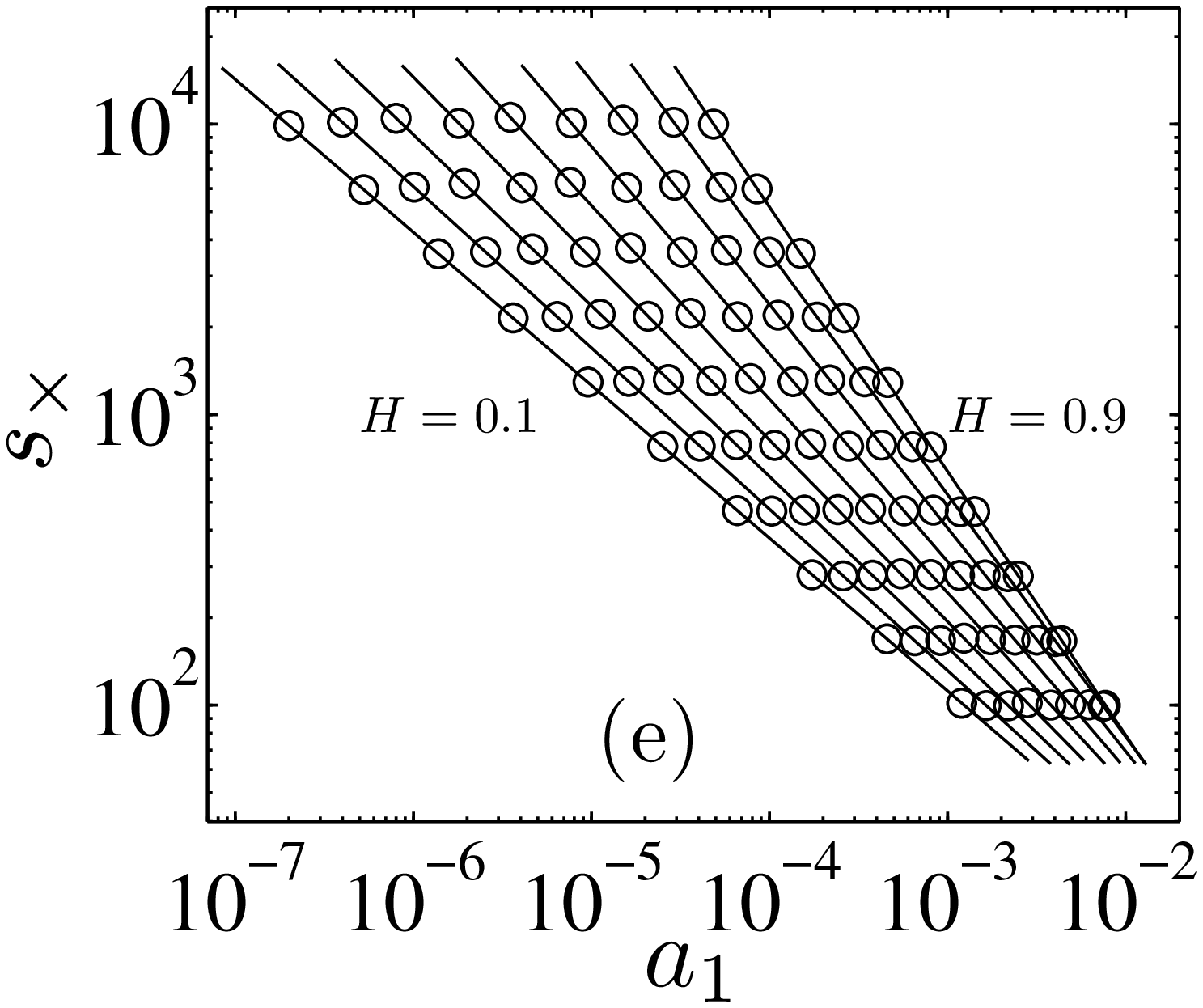}
\includegraphics[width=0.32\textwidth]{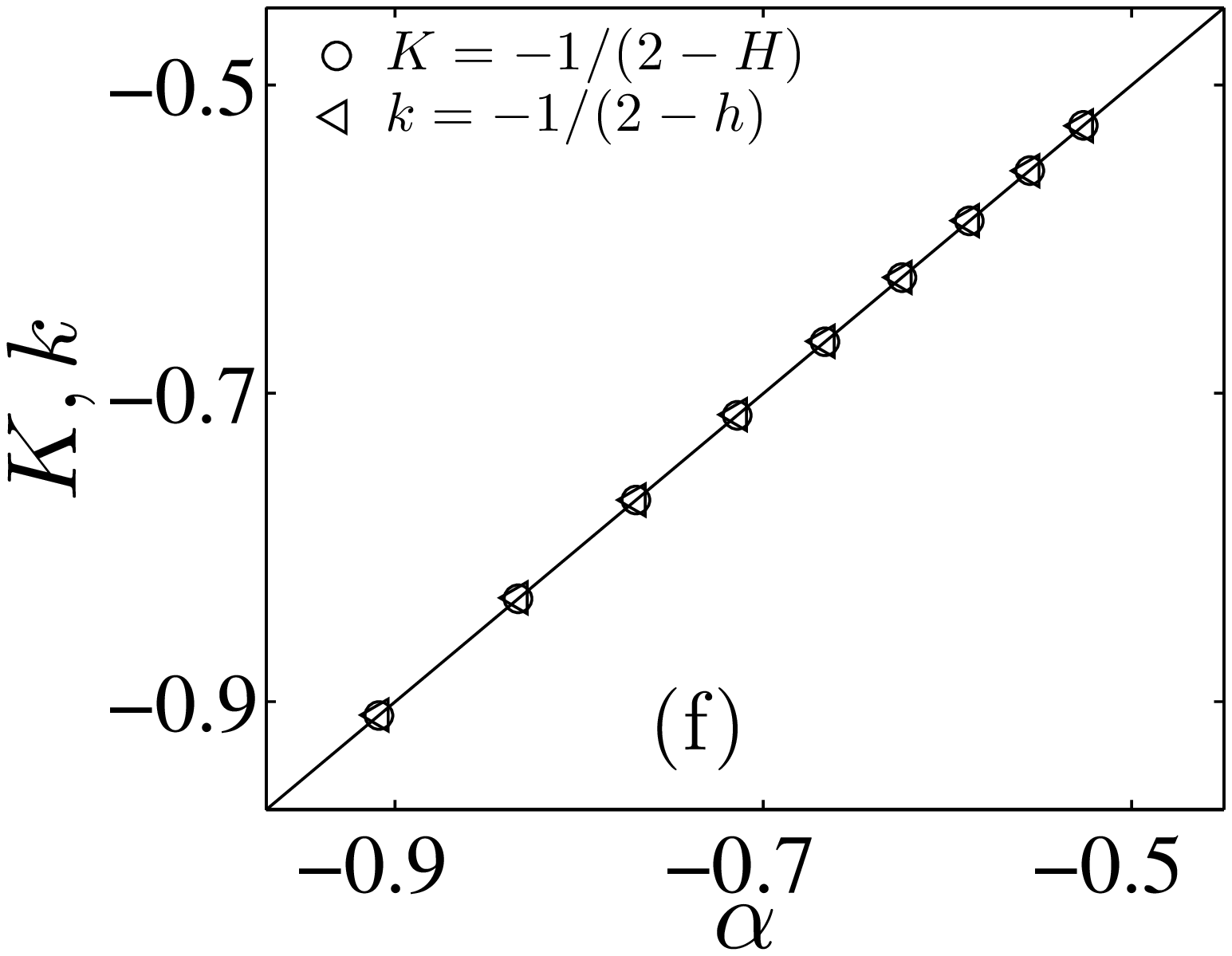}
  \caption{Effect of linear trend on the CDMA algorithm. Each curve in (a-d) represents a fluctuation function averaged over 50 repeated simulations. (a) Log-log plots of $\langle F \rangle$ of the FGNs with $H=0.1$ contaminated by a linear trend with $a_1=9.55\times10^{-6}$ in the increments. (b) Log-log plots of $\langle F \rangle$ of the FGNs for $H=0.3$ and $a_1=2.70\times10^{-5}$. (c) Log-log plots of $\langle F \rangle$ of the FGNs for $H=0.7$ and $a_1=2.18\times10^{-4}$. (d) Log-log plots of $\langle F \rangle$ of the FGNs for $H=0.9$ and $a_1=4.61\times10^{-4}$. (e) Power-law dependence of the crossover scale $s_{\times}$ on the linear coefficient $a_1$ for different Hurst indexes, varying from 0.1 (left) to 0.9 (right) with a step of 0.1. (f) Validation of Eq.~(\ref{Eq:p=1:CDMA:sx}). The crossover exponent $\alpha$ is the power-law exponent in (e) and $K=-1/(2-H)$ and $k=-1/(2-h)$, where $H$ is the input Hurst index for the generation of FGNs and $h$ is the estimated Hurst index of the generated FGNs using CDMA.}
\label{Fig:p=1:CDMA}
\end{figure*}

When $\theta=1$, we have $A=-{a_1(s-1)}/{2}$ and $L={a_1(s^{2}-1)}/{6}$. It follows that
\begin{equation}
\begin{aligned}
\label{Eq:p=1:FDMA:Fz}
  F_{z}^{2}& \approx F_{x}^{2}+a_1^{2}s_{1}^{2}\left[\frac{s^{2}}{36}-\frac{Ns}{12}+\frac{N^{2}}{12}\right]\\
           &\approx F_{x}^{2}+\frac{a_1^{2}N^{2}s^{2}}{12}
\end{aligned}
\end{equation}
The crossover $s_{\times}$ is derived as follows
\begin{equation}
  s_{\times}= \left(\frac{\sqrt{12}b}{a_1N}\right)^{{1}/{(1-H)}}
  \label{Eq:p=1:FDMA:sx}
\end{equation}
which depends on $a_1$ and $N$. We notice that the crossover scales for the FDMA and the BDMA have approximately the same expression.

\subsection{Numerical simulations}
\label{S2:p=1:Numerical}

We now perform numerical simulations to verify the correctness of the main results derived in Sec.~\ref{S2:p=1:Analytic}, in particular Eq.~(\ref{Eq:p=1:CDMA:sx}) for the CDMA method ($\theta=0.5$), Eq.~(\ref{Eq:p=1:BDMA:sx}) for the BDMA method ($\theta=0$) and Eq.~(\ref{Eq:p=1:FDMA:sx}) for the FDMA method ($\theta=1$). The procedures of numerical experiments are the same as for the case of constant shift in Sec.~\ref{S2:p=0:Numerical}.

The results for the CDMA method are illustrated in Fig.~\ref{Fig:p=1:CDMA}. In Fig.~\ref{Fig:p=1:CDMA}a, we show the averaged fluctuation function of the FGNs with $H=0.1$ contaminated by a linear trend with $a_1=9.55\times10^{-6}$. We observe an evident crossover $s_{\times}$ in the fluctuation function. When $s \ll s_{\times}$, the fluctuation function overlaps excellently with the fluctuation function of FGNs with the slope being $H=0.1$. When $s\gg s_{\times}$, the fluctuation function overlaps excellently with the fluctuation function of the linear trend $a_1t$ with the slope being $H=2$. These observations are consistent with Eq.~(\ref{Eq:p=1:CDMA:Fz}). We present respectively the results for $H=0.3$ and $a_1=2.70\times10^{-5}$ in Fig.~\ref{Fig:p=1:CDMA}b, for $H=0.7$ and $a_1=2.18\times10^{-4}$ in Fig.~\ref{Fig:p=1:CDMA}c, and for $H=0.9$ and $a_1=4.61\times10^{-4}$ in Fig.~\ref{Fig:p=1:CDMA}d. All these results are also consistent with the prediction of Eq.~(\ref{Eq:p=1:CDMA:Fz}).

%

We adopt the same procedure as the case of constant shift in the determination of the crossover scale $s_{\times}$ for different $H$ values. For each $H$, we choose 10 values of $a_1$, which are evenly spaced in logarithmic scales. Figure \ref{Fig:p=1:CDMA}e shows the dependence of $s_{\times}$ as a function of $a_1$ for different $H$ values. For fixed $a_1$, $s_{\times}$ increases with $H$, suggesting that stronger long-term correlation in the FGNs corresponds to wider scaling range in the intrinsic fluctuation function. For fixed $H$, $s_{\times}$ decreases with $a_1$, indicating that stronger trend will narrow the scaling range of the intrinsic FGNs and make it more difficult to determine the intrinsic Hurst index. For every $H$ value, we observe a nice power-law relationship:
\begin{equation}
  s_{\times} \sim a_1^{\alpha},
  \label{Eq:p=1:CDMA:sx:a1:alpha}
\end{equation}
which is consistent with the power-law form expressed in Eq.~(\ref{Eq:p=1:CDMA:sx}).

\begin{figure*}
\centering
\includegraphics[width=0.32\textwidth]{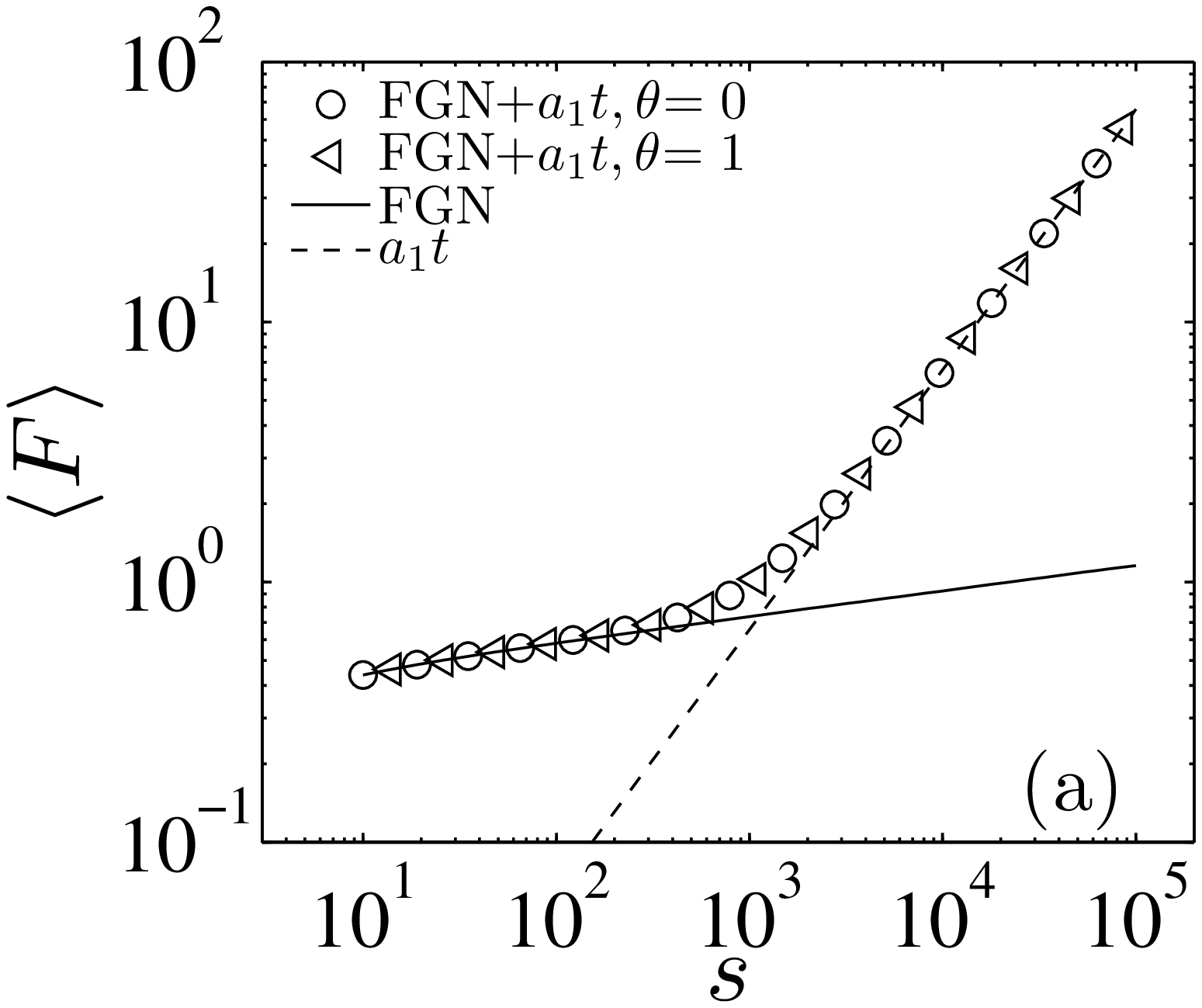}
\includegraphics[width=0.32\textwidth]{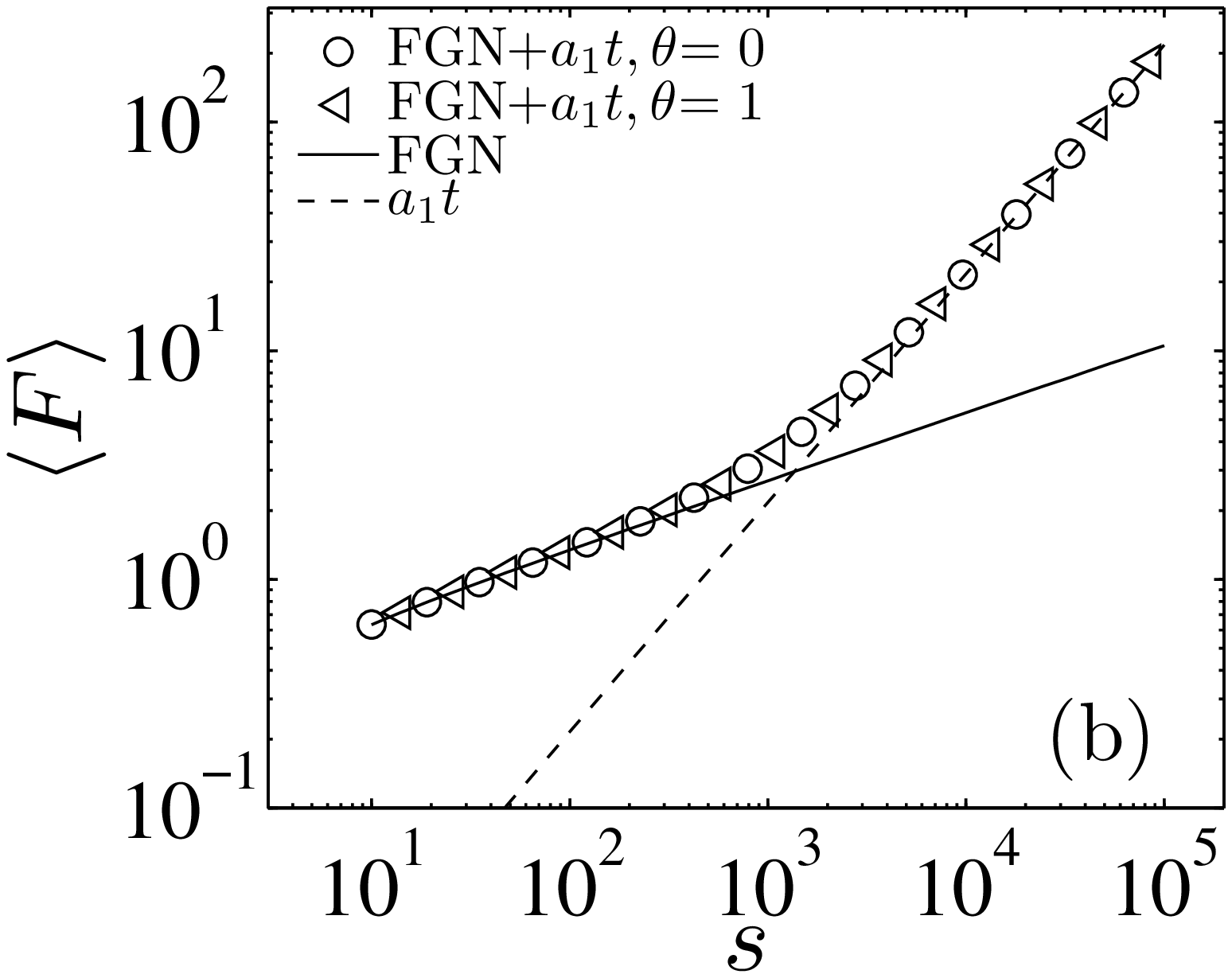}
\includegraphics[width=0.32\textwidth]{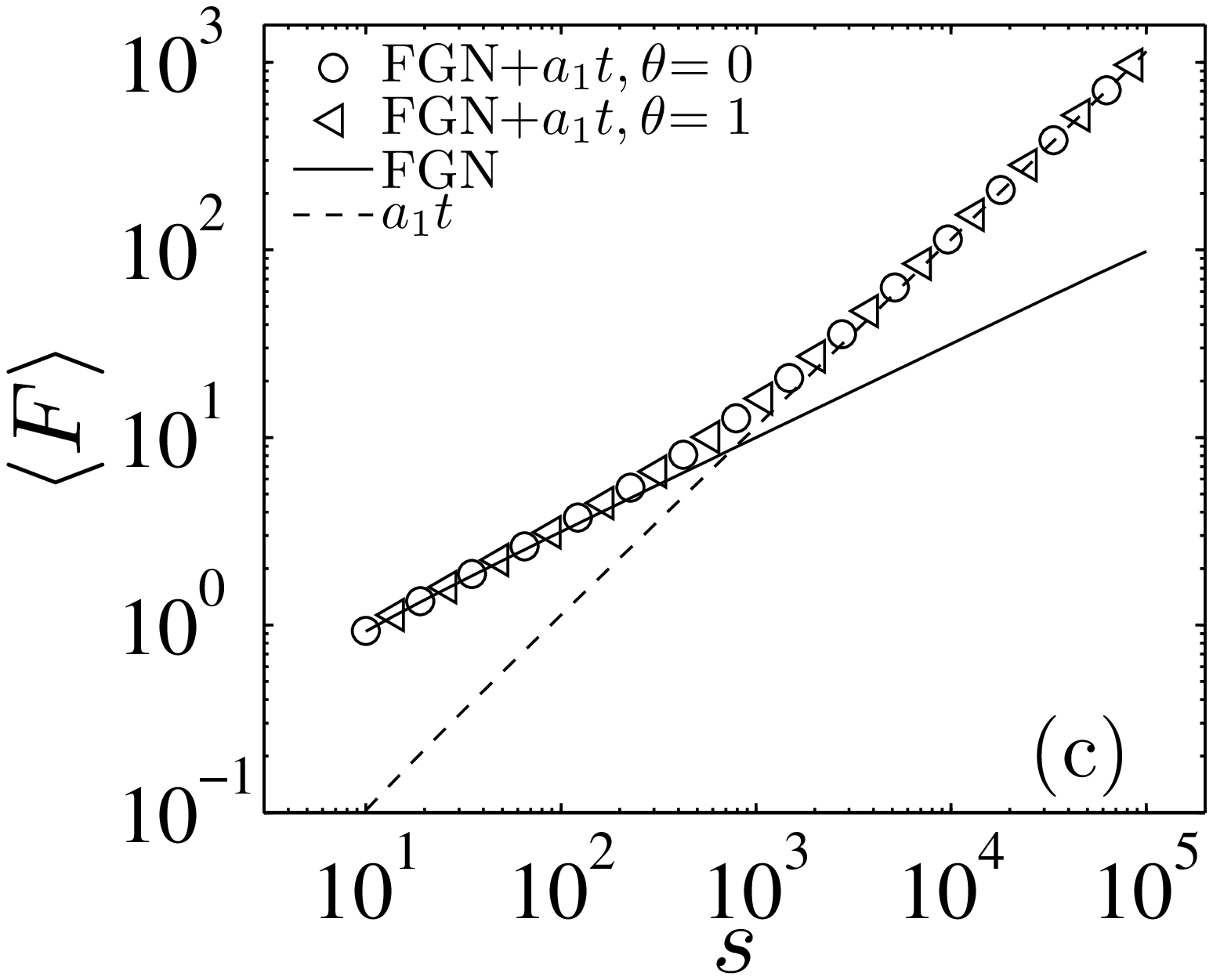}
\includegraphics[width=0.32\textwidth]{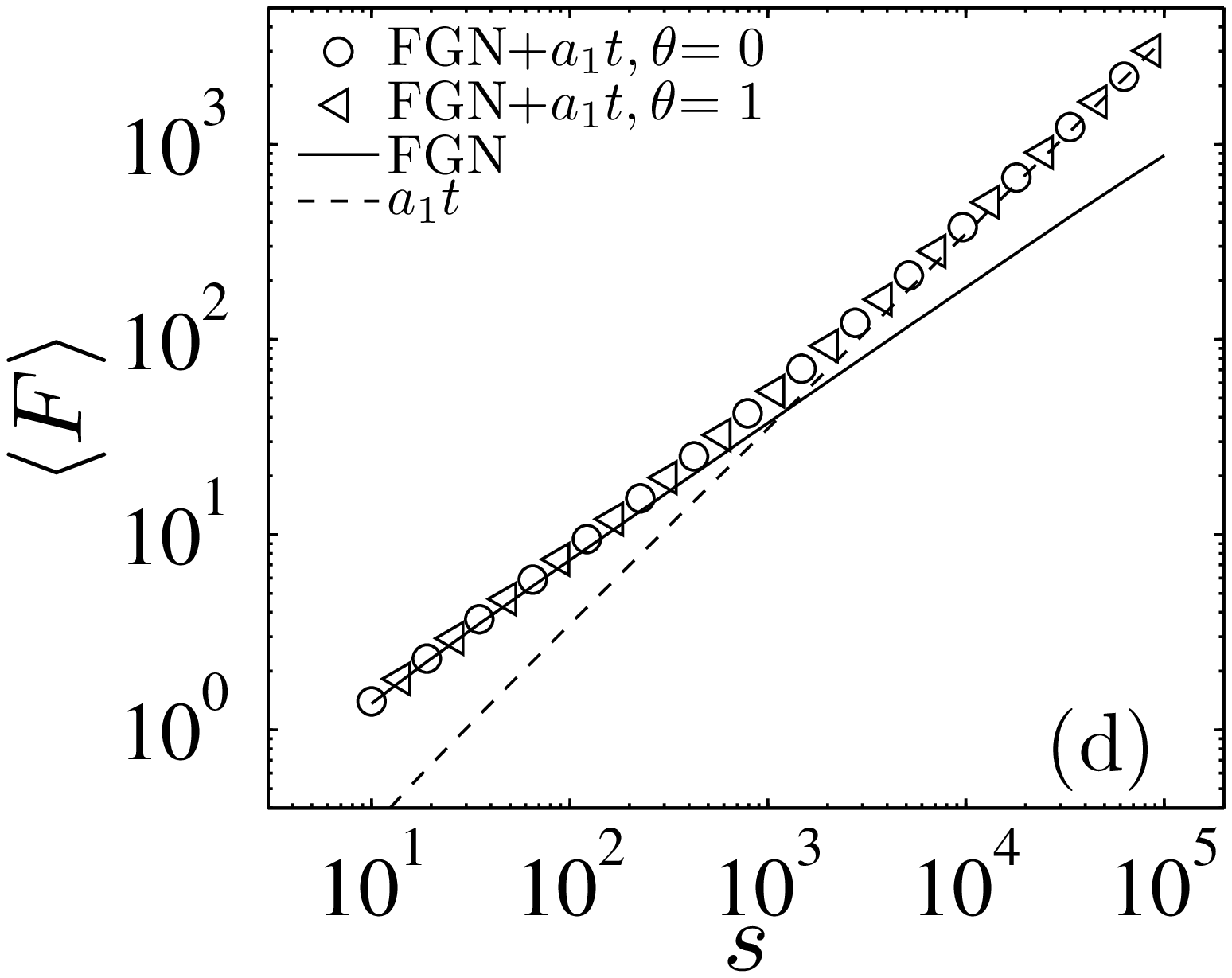}
\includegraphics[width=0.32\textwidth]{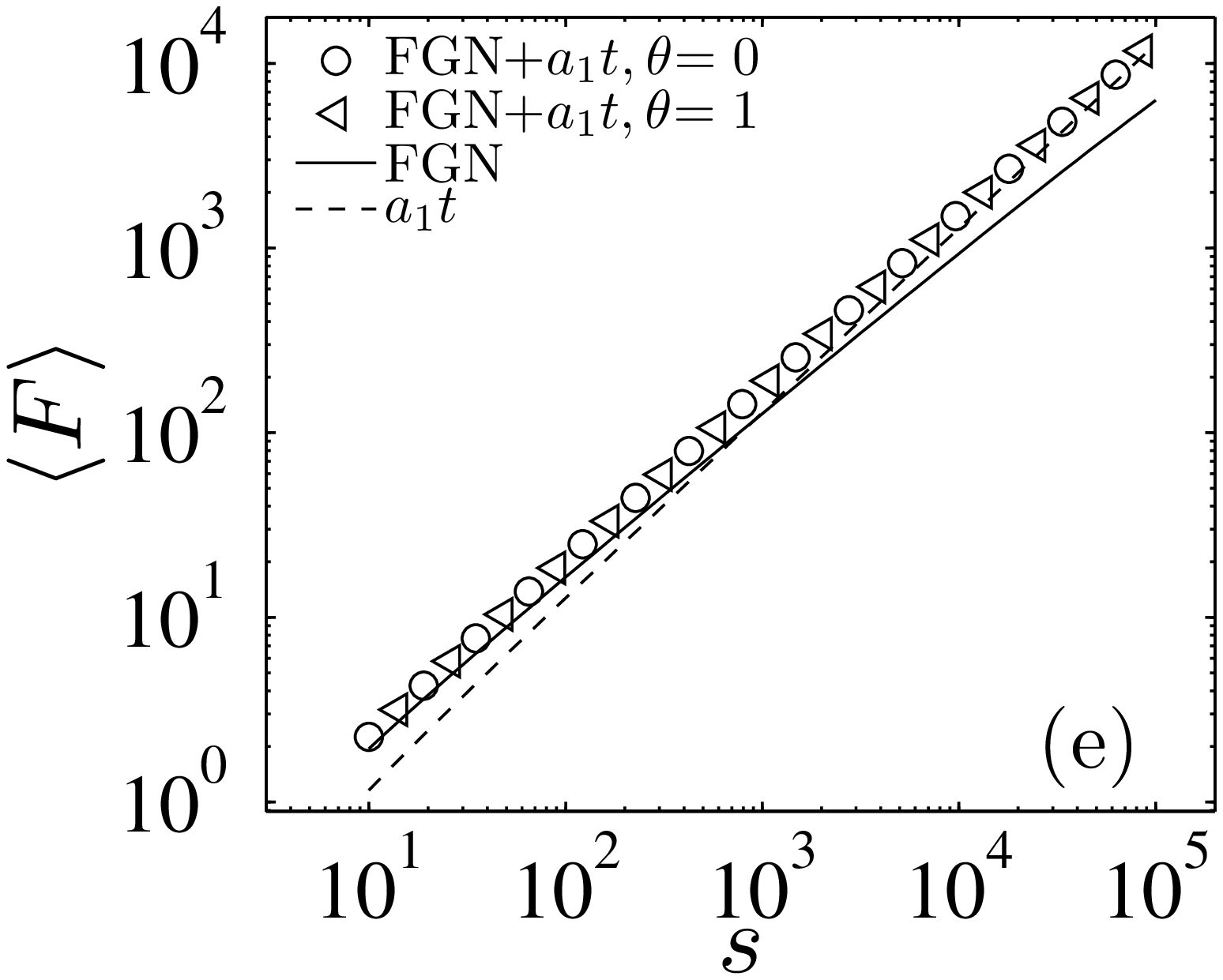}
\includegraphics[width=0.32\textwidth]{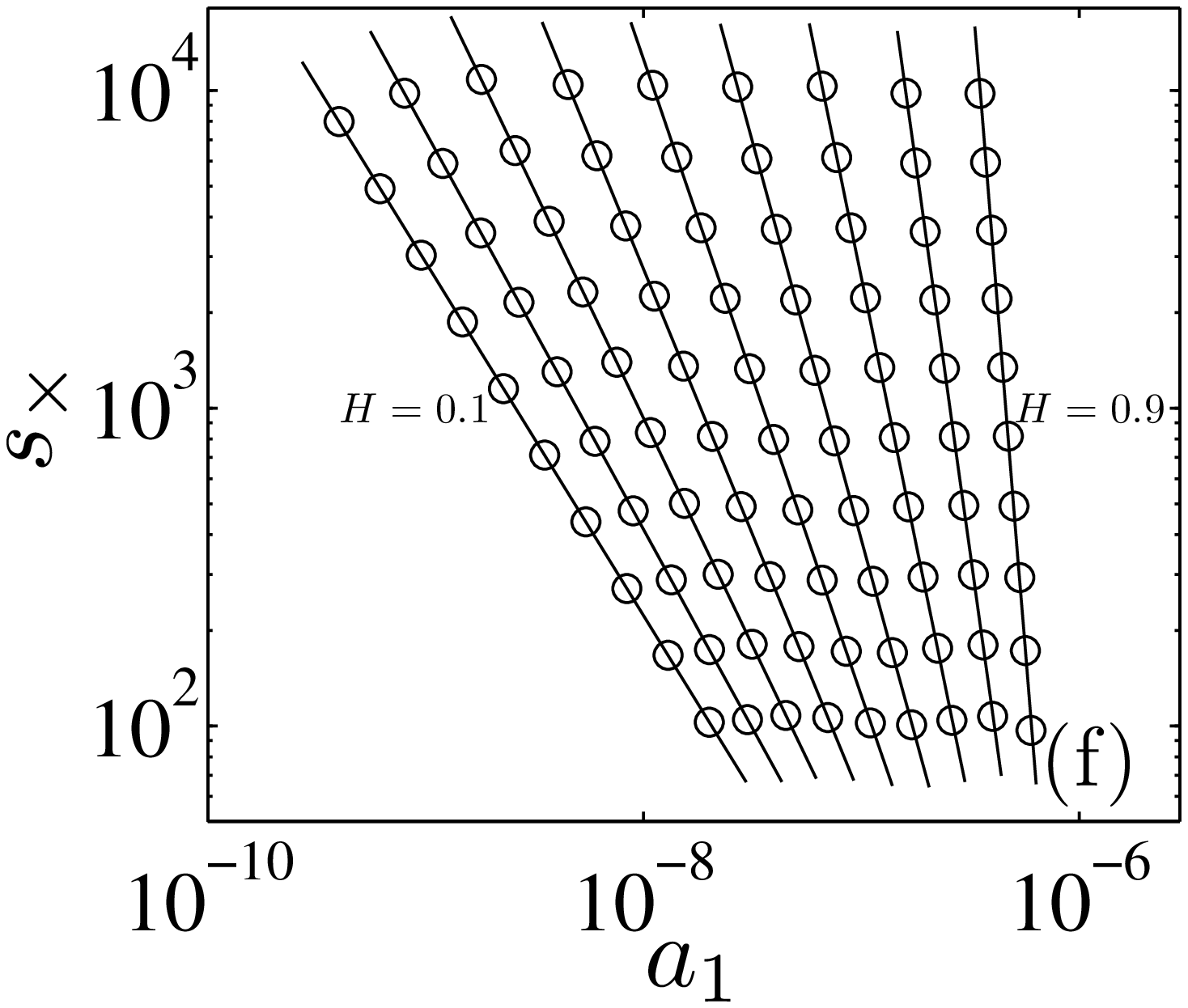}
\includegraphics[width=0.32\textwidth]{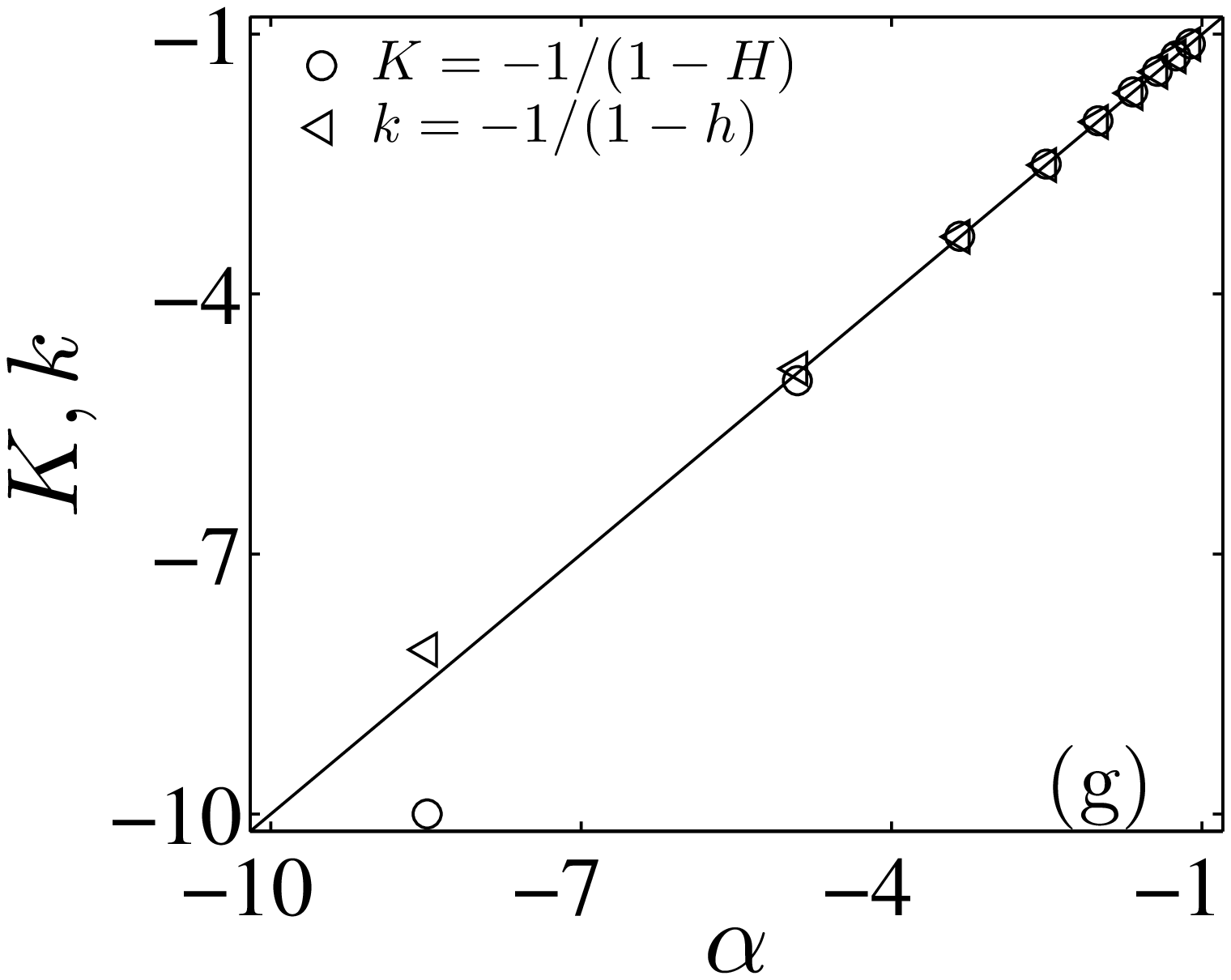}
\includegraphics[width=0.32\textwidth]{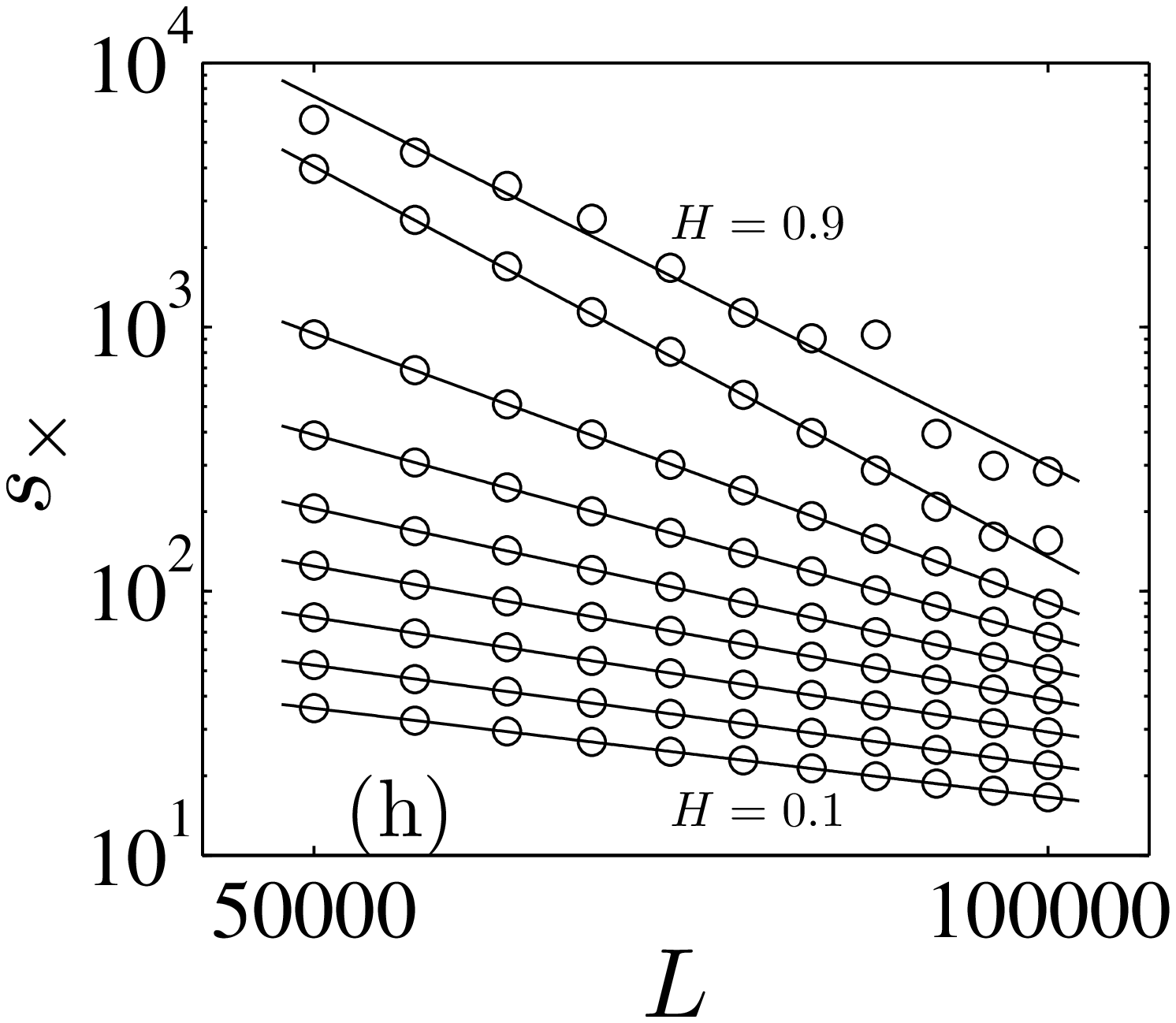}
\includegraphics[width=0.32\textwidth]{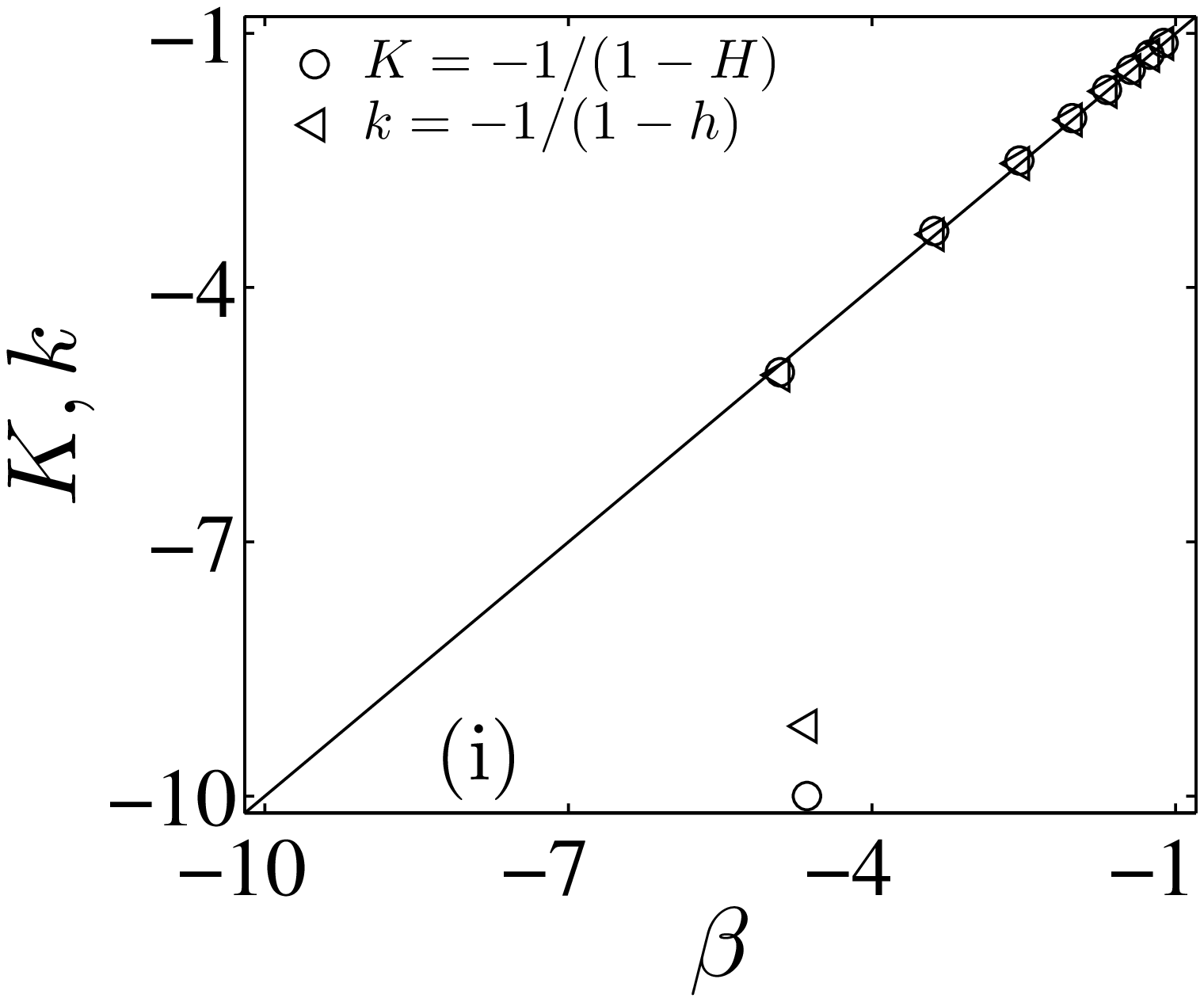}
  \caption{Effect of linear trend on the BDMA and FDMA algorithms. Each curve in (a-e) represents the fluctuation function averaged over 50 repeated simulations. (a) Log-log plots of $\langle F \rangle$ of the FGNs with $H=0.1$ contaminated by a linear trend with $a_1=2.28\times10^{-9}$ in the increments. (b) Log-log plots of $\langle F \rangle$ against $s$ for the case of $H=0.3$ and $a_1=7.53\times10^{-9}$. (c) Log-log plots of $\langle F \rangle$ against $s$ for the case of $H=0.7$ and $a_1=3.95\times10^{-8}$. (d) Log-log plots of $\langle F \rangle$ against $s$ for the case of $H=0.7$ and $a_1=1.21\times10^{-7}$. (e) Log-log plots of $\langle F \rangle$ against for the case of $H=0.9$ and $a_1=4.45\times10^{-7}$. (f) Power-law dependence of the crossover scale $s_{\times}$ obtained from the BDMA method on the linear coefficient $a_1$ for different Hurst indexes, varying from 0.1 (left) to 0.9 (right) with a step of 0.1. (g) Validation of $s_{\times} \approx a_1^{-1/(1-H)}$ in Eq.~(\ref{Eq:p=1:BDMA:sx}). The crossover exponent $\alpha$ is the power-law exponent in (f) and $K=-1/(1-H)$ and $k=-1/(1-h)$, where $H$ is the input Hurst index for the generation of FGNs and $h$ is the estimated Hurst index of the generated FGNs using BDMA. (h) Power-law dependence of the crossover scale $s_{\times}$ obtained from the BDMA method on the time series length $N$ for different Hurst indexes, varying from 0.1 (bottom) to 0.9 (top) with a step of 0.1. (i) Validation of $s_{\times} \approx N^{-1/(1-H)}$ in Eq.~(\ref{Eq:p=1:BDMA:sx}). The crossover exponent $\beta$ is the power-law exponent in (h) and $K=-1/(1-H)$ and $k=-1/(1-h)$. }
\label{Fig:p=1:BDMA:FDMA}
\end{figure*}


Figure \ref{Fig:p=1:CDMA}f shows that the lines become steeper for larger Hurst indexes. We fit the data points for each $H$ to estimate the power-law exponent $\alpha$. We then define and calculate the following two quantities:
\begin{equation}
  K=-1/(2-H)
\end{equation}
and
\begin{equation}
  k=-1/(2-h),
\end{equation}
where $H$ is the input Hurst indexes for the synthesis of the FGNs and $h$ is the output Hurst indexes of the synthesized FGNs using the CDMA method. We plot $K$ against $\alpha$ and $k$ against $\alpha$ in Fig.~\ref{Fig:DMA:ConstantShift}f. We observe that all the points fall on the diagonal line
\begin{equation}
  K\approx k\approx\alpha.
  \label{Eq:p=1:CDMA:sx:K:k:alpha}
\end{equation}
Equations (\ref{Eq:p=1:CDMA:sx:a1:alpha}) and (\ref{Eq:p=1:CDMA:sx:K:k:alpha}) verify excellently Eq.~(\ref{Eq:p=1:CDMA:sx}).

%

The results for the BDMA and FDMA methods are depicted in Fig.~\ref{Fig:p=1:BDMA:FDMA}. Applying the BDMA and FDMA methods, we show in Fig.~\ref{Fig:p=1:BDMA:FDMA}a the averaged fluctuation functions of the FGNs with $H=0.1$ contaminated by a linear trend with $a_1=2.28\times10^{-9}$. The two curves overlap nicely. We observe an evident crossover $s_{\times}$ in the fluctuation functions. When $s \ll s_{\times}$, the fluctuation function overlaps excellently with the fluctuation function of FGNs with the Hurst index $H=0.1$, which is shown as a solid straight line. When $s\gg s_{\times}$, the fluctuation function overlaps excellently with the fluctuation function of the linear trend $a_1t$ with the slope being $H=1$, which is shown as a dashed straight line. We present respectively the results for $H=0.3$ and $a_1=7.53\times10^{-9}$ in Fig.~\ref{Fig:p=1:BDMA:FDMA}b, $H=0.7$ and $a_1=3.95\times10^{-8}$ in Fig.~\ref{Fig:p=1:BDMA:FDMA}c, for $H=0.7$ and $a_1=1.21\times10^{-7}$ in Fig.~\ref{Fig:p=1:BDMA:FDMA}d, and for $H=0.9$ and $a_1=4.45\times10^{-7}$ in Fig.~\ref{Fig:p=1:BDMA:FDMA}e. All these results are consistent with the prediction of Eq.~(\ref{Eq:p=1:BDMA:Fz}) and Eq.~(\ref{Eq:p=1:FDMA:Fz}), which have the same expression. Note that these two expressions are approximations of Eq.~(\ref{Eq:p=1:DMA:Fz}). The difference between the $F_z$ functions of the BDMA and FDMA methods is about $2AL$. These results show that this difference is ignorable and the approximation leading to Eq.~(\ref{Eq:p=1:BDMA:Fz}) and Eq.~(\ref{Eq:p=1:FDMA:Fz}) is reasonable.

We determine the crossover scale $s_{\times}$ for different $H$ and $a_1$ values. Figure \ref{Fig:p=1:BDMA:FDMA}f shows the dependence of $s_{\times}$ as a function of $a_1$ for different $H$ values. It is again found that $s_{\times}$ increases with $H$ for fixed $a_1$ and decreases with $a_1$ for fixed $H$. For every $H$ value, we observe a nice power-law relationship:
\begin{equation}
  s_{\times} \sim a_1^{\alpha}.
  \label{Eq:p=1:BDMA:sx:a1:alpha}
\end{equation}
We determine the power-law exponents $\alpha$ for different $H$ values and plot $K=-1/(1-H)$ against $\alpha$ and $k=-1/(1-h)$ against $\alpha$ in Fig.~\ref{Fig:p=1:BDMA:FDMA}g. We observe that all the points fall on the diagonal line
\begin{equation}
  K\approx k\approx\alpha,
  \label{Eq:p=1:BDMA:sx:K:k:alpha}
\end{equation}
except for $H=0.9$. Equations (\ref{Eq:p=1:BDMA:sx:a1:alpha}) and (\ref{Eq:p=1:BDMA:sx:K:k:alpha}) verify excellently Eq.~(\ref{Eq:p=1:BDMA:sx}).

We now investigate the dependence of the crossover scale $s_{\times}$ on the length $N$ of time series. In Fig.~\ref{Fig:p=1:BDMA:FDMA}h, we plot $s_{\times}$ against $N$ in log-log scales for different Hurst indexes $H$ and $a_1$ values: $a_1=7\times10^{-8}$ for $H=0.1$, $a_1=2\times10^{-7}$ for $H=0.2$, $a_1=4\times10^{-7}$ for $H=0.3$, $a_1=6\times10^{-7}$ for $H=0.4$, $a_1=1\times10^{-6}$ for $H=0.5$, $a_1=2\times10^{-6}$ for $H=0.6$, $a_1=3\times10^{-6}$ for $H=0.7$, $a_1=5.4\times10^{-6}$ for $H=0.8$, $a_1=6.4\times10^{-6}$ for $H=0.9$. In our numerical experiments, $N$ ranges from 50000 to 100000 with a step of 5000. The determination of this range is not arbitrary. If we include shorter time series, say $N\sim10^4$, we have $s_\times \sim 10^{3.3}$, which is larger than $N/10$ and thus cannot be detected in the fluctuation function. If we include longer time series, say $N\sim10^6$, we have $s_\times \sim 10^{1\sim2}$, which is again hard to identify.

For every $H$ value, we observe a nice power-law relationship:
\begin{equation}
  s_{\times} \sim N^{\beta}.
  \label{Eq:p=1:BDMA:sx:N:beta}
\end{equation}
The power-law scaling for $H=0.9$ is the worst. When $N$ is small (left part) or large (right part), the crossover scale is large or small, which makes it very difficult to identify because the crossovers are near the end points of the fluctuation function.

We determine the power-law exponents $\beta$ for different $H$ values and plot $K=-1/(1-H)$ against $\beta$ and $k=-1/(1-h)$ against $\beta$ in Fig.~\ref{Fig:p=1:BDMA:FDMA}i. We observe that all the points fall on the diagonal line
\begin{equation}
  K\approx k\approx\beta,
  \label{Eq:p=1:BDMA:sx:K:k:beta}
\end{equation}
except for $H=0.9$. Equations (\ref{Eq:p=1:BDMA:sx:N:beta}) and (\ref{Eq:p=1:BDMA:sx:K:k:beta}) verify excellently Eq.~(\ref{Eq:p=1:BDMA:sx}).

%
%
%
%
%

\section{Summary}

In this paper, using fractional Gaussian noises (FGNs) with different Hurst indexes, we have investigated the effects of polynomial trends on the scaling behaviors and the performance of three widely used DMA methods including backward algorithm (BDMA), centered algorithm (CDMA) and forward algorithm (FDMA). We derived a general framework for polynomial trends and obtained analytical results for constant shifts and linear trends in the FGNs. We performed extensive numerical experiments which confirm excellently the analytical derivations.

We first considered constant shifts in the FGNs. We found that the behavior of the CDMA method is not influenced by constant shifts. In contrast, constant shifts result in a crossover $s_{\times}$ in the fluctuation function when the BDMA and FDMA methods are applied. The crossover $s_{\times}$ scales as a power law of the constant shift, which increases with the Hurst index $H$ and decreases with the strength of constant shift $a_0$.

We then considered linear trends in the FGNs. We found that a linear trend $a_1t$ causes a crossover in the fluctuation function of the CDMA method. The crossover scale $s_{\times}$ is a power law of the strength $a_1$ of the linear trend, which increases with the Hurst index $H$ and decreases with the strength of linear trend $a_1$. A linear trend also results in a crossover in the fluctuation function when the BDMA and FDMA methods are applied. The crossover $s_{\times}$ scales as a power law of the production of the strength of linear trend and the length of the time series, which increases with the Hurst index $H$ and decreases with the strength of constant shift $a_1$ and the length of linear trend. It is intriguing that longer time series are less resistent to linear trends when the BDMA and FDMA methods are adopted.

When a crossover appears, the left part of the fluctuation function with the scales less than $s_{\times}$ reflects the behavior of FGNs, while the right part with the scales greater than $s_{\times}$ is dominated by the polynomial trend. Our findings show that time series with larger Hurst indexes are more resistent to polynomial trends and the polynomial trend has the same effect on the BDMA and the FDMA. Because a large crossover scale will make it easier to estimate the intrinsic Hurst index, we conclude that the CDMA method outperforms the BDMA and FDMA methods in the presence of polynomial trends (see Fig.~\ref{Fig:DMA:ConstantShift}a versus Fig.~\ref{Fig:DMA:ConstantShift}e for constant shifts and Fig.~\ref{Fig:p=1:CDMA}e versus Fig.~\ref{Fig:p=1:BDMA:FDMA}f for linear trends).

\begin{acknowledgments}
  We acknowledge financial support from the National Natural Science Foundation of China under grant no. 11375064 and the Fundamental Research Funds for the Central Universities.
\end{acknowledgments}

\bibliographystyle{naturemag}
\bibliography{E:/Papers/Auxiliary/Bibliography}

\end{document}